\documentclass[aps,pra,reprint,superscriptaddress,linenumbers,secnumarabic]{revtex4-2}

\usepackage{graphicx}
\usepackage{dcolumn}
\usepackage{bm}

\usepackage{placeins}

\usepackage{multibib}
\usepackage{lineno}
\usepackage{braket}
\usepackage{amsmath}
\usepackage{bibentry}

\usepackage{xcolor} 

\graphicspath{{figures/}}

\colorlet{myblue}{red!80!black}

\newcommand\hl[1]{%
  \bgroup
  \hskip0pt\color{red!80!black}%
  #1%
  \egroup
}
\newcommand\hlsyc[1]{%
	\bgroup
	\hskip0pt\color{red!80!black}%
	#1%
	\egroup
}

\newcommand{\Renyi}{R\'{e}nyi\hspace{1mm}}
\newcommand{\nocontentsline}[3]{}
\newcommand{\tocless}[3]{\bgroup\let\addcontentsline=\nocontentsline#1{#3}\egroup}

\linenumbers

\begin{document}

\title{Quantifying Quantum Computational Advantage on a Processor of Ultracold Atoms}

\author{Yong-Guang Zheng}

\thanks{Y.-G.Z., Y.-C.S. and W.-Y.Z. contributed equally to this work.}


\author{Ying-Chao Shen}
\thanks{Y.-G.Z., Y.-C.S. and W.-Y.Z. contributed equally to this work.}

\author{Wei-Yong Zhang}

\thanks{Y.-G.Z., Y.-C.S. and W.-Y.Z. contributed equally to this work.}

\author{An Luo}

\author{Ying Liu}

\author{Ming-Gen He}

\author{Hao-Ran Zhang}

\author{Wan Lin}

\author{Han-Yi Wang}

\author{Zi-Hang Zhu}


\author{Pei-Yue Qiu}
\author{Tian-Yi Wang}

\author{Ming-Cheng Chen}

\affiliation{Hefei National Research Center for Physical Sciences at the Microscale and School of Physical Sciences, University of Science and Technology of China, Hefei 230026, China}
\affiliation{CAS Center for Excellence in Quantum Information and Quantum Physics, University of Science and Technology of China, Hefei 230026, China}

\author{Chao-Yang Lu}

\affiliation{Hefei National Research Center for Physical Sciences at the Microscale and School of Physical Sciences, University of Science and Technology of China, Hefei 230026, China}
\affiliation{CAS Center for Excellence in Quantum Information and Quantum Physics, University of Science and Technology of China, Hefei 230026, China}
\affiliation{Hefei National Laboratory, University of Science and Technology of China, Hefei 230088, China}

\author{Supanut Thanasilp}

\affiliation{Centre for Quantum Technologies, National University of Singapore, 3 Science Drive 2, Singapore 117543}
\affiliation{Chula Intelligent and Complex Systems Center of Excellence, Department of Physics, Faculty of Science, Chulalongkorn University, Bangkok, Thailand, 10330}
\affiliation{Siam Quantum Square, Faculty of Science, Chulalongkorn University, Bangkok, Thailand}

\author{Dimitris G. Angelakis}
\affiliation{Centre for Quantum Technologies, National University of Singapore, 3 Science Drive 2, Singapore 117543}
\affiliation{Institute for Quantum Computing and Quantum Technologies, NCSR Demokritos, Greece}
\affiliation{AngelQ Quantum Computing, 531A Upper Cross Street \#04-95 Hong Lim Complex, Singapore 051531}
\affiliation{School of Electronic and Computer Science, University of Southampton, University Road, Southampton, SO17 1BJ, United Kingdom}

\author{Zhen-Sheng Yuan}

\author{Jian-Wei Pan}
\affiliation{Hefei National Research Center for Physical Sciences at the Microscale and School of Physical Sciences, University of Science and Technology of China, Hefei 230026, China}
\affiliation{CAS Center for Excellence in Quantum Information and Quantum Physics, University of Science and Technology of China, Hefei 230026, China}
\affiliation{Hefei National Laboratory, University of Science and Technology of China, Hefei 230088, China}

\date{\today}

\begin{abstract}

Nonequilibrium dynamics of quantum many-body systems is challenging for classical computing, providing opportunities for demonstrating practical quantum computational advantage with analogue quantum simulators.
Owing to the intimate connection with a random matrix ensemble, it is proposed to be classically intractable to sample the driven thermalized many-body states of a Bose-Hubbard system, and further extract multi-point correlations from the output-strings for characterizing quantum systems.
Here, leveraging dedicated precise manipulations and atom-number-resolved detection through a quantum gas microscope with bichromatic superlattices, we perform sampling of the driven Hubbard chains and two-leg ladders in the thermalized phase involving up to 64 sites with 20 atoms, yielding a Hilbert space dimension of $10^{19}$ and outpacing the most powerful supercomputer in terms of sampling rate by three orders of magnitude. 
The volume law scaling of the \Renyi entanglement entropy in the thermalized phase is observed, which hinders efficient classical simulation for large systems. 
We employ the Bayesian tests to verify that our prepared systems operate in the driven thermalized phase. Multi-point correlations of up to 14th-order extracted from the experimental samples offer clear distinctions between the thermalized and many-body-localized phases, where classical computations such as tensor network fails to give accurate and faithful predictions within a reasonable time cost. 
Our work demonstrates the sampling of a interacting chaotic system performed on a quantum processor of ultracold atoms and opens the door of utilizable quantum computational advantage in simulating Floquet dynamics of many-body systems.

\end{abstract}

\maketitle

\section{Introduction}
Periodically driven systems play crucial roles in exotic quantum phases of matter, such as hosting discrete time crystals \cite{Zhang2017e,Choi2017,Mi2021} and prethermalization \cite{Rubio-Abadal2020} or stablising many-body scars \cite{Bluvstein2021,Su2022}.
In the driven ergodic phases, emergent long-range interactions lead to an infinite temperature thermalization \cite{Thanasilp2020,DAlessio2014,Lazarides2014,Ponte2015} and a chaotic spreading throughout the entire Hilbert space. 
Rapidly increasing entanglement during the evolution hinders efficient classical simulations of these dynamics with the currently known algorithms \cite{Schuch2008,Daley2022}, and in turn, impedes extracting multi-point correlations to characterize the quantum many-body system \cite{Schweigler2017}.

Quantum computation is expected to prevail over its classical counterpart in simulating highly entangled quantum systems \cite{Feynman1982, Preskill2012}. 
However, building fault-tolerant digital universal quantum computers remains challenging due to the fragility of qubits, the need for extensive error correction, and the substantial overhead in terms of qubits and gates required \cite{Terhal2015,Daley2022,Flannigan2022,Kim2023}. 
On the other hand, the rapid development of analogue quantum simulators makes it a promising alternative for practical applications of near-term quantum hardware \cite{Georgescu2014,Altman2019,Guo2024,Shaw2024,Young2024,King2024}. 
Here, quantum simulations of various physics models are considered as utilizable applications of quantum computers \cite{Daley2022,Kim2023}.
Analogue simulators with the inherent many-body nature directly mimic the problems of interest in condensed matter and high-energy physics and, in general, are more hardware-friendly than digital ones, as they require much less local controls and the extra layer of translation into gates is not necessary \cite{Flannigan2022,Daley2022,Trivedi2022}.

\begin{figure*}[!thb]
    \includegraphics[width=\linewidth]{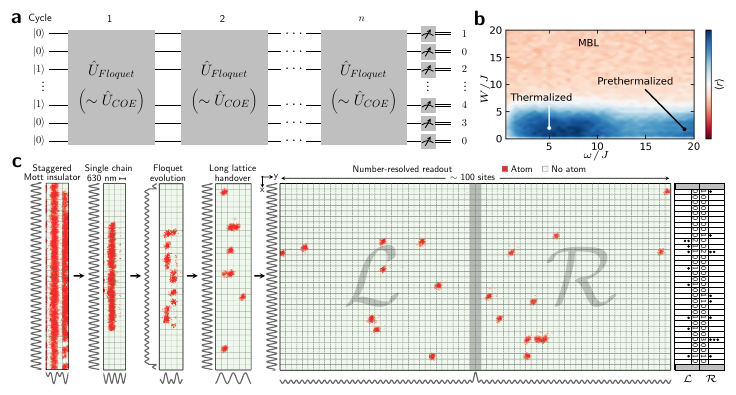}%
    \caption{\label{fig:Setup}
    Sampling the Floquet many-body systems on a quantum processor of ultracold atoms. 
    (a) Quantum circuit illustration of the experiment. 
    Starting from a product state with $N_b$ bosons sitting in a 1D $L$-site chain or a two-leg ladder with $L \times 2$ sites, we sample the output of the final states on the Fock basis after several cycles of the Floquet unitary evolution. 
    (b) The phase diagram of a driven Bose-Hubbard chain. The colorbar shows the expectation value of the level spacing $\braket{r}$. For low driving frequency $\omega$ and weak disorders $W$, the system will thermalize after several cycles of driving. If the intensity of the disorder increases, the thermalization breaks down and the system arrives at a MBL phase. In this work, we focus on the thermalized and MBL phases. We get this phase diagram from a system of 10 sites with 6 particles. Both axes are in unit of tunneling $J/h=166$ Hz.
    (c) Sketch of the sampling experiments in a two-leg ladder system. We implement the protocol in our analogue quantum simulator based on optical lattices with a quantum gas microscope. First, a defect-free Mott insulator is prepared via staggered cooling. Utilizing the site-resolved addressing, we trim an $L \times 2$-site two-leg ladder with $N_b$ particles in the middle of its left leg. Then, the system evolves under a periodic-driven Bose-Hubbard model to the final states in the presence of a disorder potential. The edges of the chain and ladder are fixed by a box trap imposed by a DMD. For the detection, we freeze the atoms by ramping up short lattices in $x$ and $y$ directions and hand over them to the long lattice in the transverse direction. Then, we ramp up a barrier between the two legs of the ladder to avoid mixing the atoms from the left and right legs during the expansion. The expansion of the atoms circumvents the pair-wise loss of multiple atoms in a single site during fluorescence imaging. 
     } 
\end{figure*}

Notably, the computational task of sampling from a periodically driven thermalized many-body system has been proposed to demonstrate utilizable quantum computational advantage backed by complexity-theoretic foundations \cite{Tangpanitanon2020,Thanasilp2020}, which is similar to cases of random circuit sampling and boson sampling and thereby is far more convincing to demonstrate advantage of quantum computers than previous works of analog quantum simulation \cite{Harrow2017}. It also gives identifications the dynamical phases in the driven system through multi-point density correlations simultaneously. 
For low-frequency driving and weak disorders, the system will thermalize, and the associated temperature will become infinite. In this limit, the Floquet operator $U_F$ is intimately related to a random matrix drawn from the circular orthogonal ensemble (COE), where sampling the probability distribution of the final state is classically intractable \cite{Tangpanitanon2020,Thanasilp2020}. In contrast, classical simulation is feasible in the many-body-localized (MBL) phase due to an area-law scaling of entanglement. 
Therefore, sampling the many-body system and extracting multi-point correlations from the samples while varying the parameters will give information on the dynamical phase diagram, which is beyond the characterizations of the quantum machine itself \cite{Arute2019,Zhong2020,Zhong2021a,Wu2021,Zhu2021a,Madsen2022,Hangleiter2023}.

In the present work, we implement a driven Bose-Hubbard model in optical lattices and draw samples from the output probability distribution encoded in the states of the thermalized phase, involving a two-leg ladder system of 64 sites with 20 particles, which is in the intractable regime for classical computation. 
Site-resolved atom addressing and pattern-programable potential engineering enable deterministic initialisation and long-time periodic driving of the system.
The stability of the set-up and precise calibrations offer us the ability to  quantitatively and reliably simulate many-body dynamics. 
After the final state is projected on a Fock basis, we utilise the atom-number readout to obtain the output strings. 
For small one-dimensional (1D) chain systems with up to 20 sites, which can still be simulated and verified classically, we validate the samples using either classical fidelity \cite{Spring2013,Crespi2013,Tillmann2013,Carolan2014,Wang2017c,Zhong2018} or Bayesian hypothesis tests \cite{Bentivegna2014,Wang2017c,Zhong2021a,Madsen2022}. 
The results of Bayesian hypothesis tests are then used to estimate the complexity of sampling the thermalized phase, providing a basis for testing scenarios that are classically intractable. 
In the two-leg ladder of 64 sites with 20 atoms, the time cost is estimated to be at least 8 days to generate a single exact sample for the \textit{Frontier} supercomputer (8,730,112 cores, 9.2 petabytes), the currently most powerful supercomputer worldwide, with currently known best algorithms and supposing its random access memory (RAM) is sufficient (the required RAM is much larger than that the \textit{Frontier} equipped, see Appendices). 
For comparison, it takes only 500 seconds to perform the same task in our quantum machine, yielding a  quantum speedup of 3 orders of magnitude. 
The quantum advantage of the sampling experiments achieved here is further applied to distinguish the thermalized phase from the MBL phase. 
The multi-point correlation functions are extracted from those samples after periodic driving and enhanced high-order correlations are observed in the driven thermalized phase compared to the MBL phase. 
The approximate algorithms of matrix product states (MPS) also break down to reproduce the results in the driven thermalized phase. 
Moreover, we manage to measure the second-order \Renyi entropy, showing a volume law of the entanglement in the thermalized phase. 
These observations provide distinctions between the thermalized phase versus the MBL phase.

\section{Experimental set-up}
Our experiments start with a two-dimensional Bose--Einstein condensate of ${}^{87}$Rb atoms, which resides in a single anti-node of the vertical lattice ($z$ lattice). Before ramping up the horizontal short lattices, we superimpose a staggered potential in the $y$ direction by introducing a long lattice, whose wavelength is double that of the short lattice \cite{Li2021}. Then, we ramp up the $xy$ short lattices to drive the phase transition to the Mott insulator. During the ramping, the entropy redistribution induced by the staggered potential renders the Mott insulator nearly defect-free (Fig. \ref{fig:Setup}(c)) \cite{Yang2020,Zhang2023}. A site-resolved addressing beam is projected to the atoms through the objective by a digital micromirror device (DMD) \cite{Zheng2022}. We select a single chain along the $x$ direction with a definite atom number $N_b$, and the rest of the atoms are pushed out by a resonant laser pulse. 

Then, we ramp up the $y$ long lattice to isolate the double wells, forming a two-leg ladder system. For the 1D chains this step is skipped. Afterwards, we drop the $x$ short lattice depth to 2.9$E_{rS}$ to initialise the dynamics, where $E_{rS}=h^2/(8 m a_S^2)$ is the recoil energy of the short lattice, $m$ is the atomic mass of ${}^{87}$Rb and $a_S=$ 630 nm denotes the lattice constant of the short lattice. The length of the chain or ladder $L$ is defined by imposing a box trap in the $x$ direction (Appendices). Throughout this work, we use the notation in the parentheses ($L$,$N_b$) denoting an $L$-site chain with $N_b$ bosons.

To periodically drive the system, the depth of the $y$ short lattice is modulated following a sinusoidal curve $V_y(t)=V_y(1+V_s\sin{2\pi \omega t})$\cite{Rubio-Abadal2020}, where $V_y=47 E_{rS}$ is the depth of the $y$ short lattice and $V_s=0.2$ is the amplitude of the modulation. At the given low depth of the $x$ short lattice, the system can be described by the non-standard Bose-Hubbard model (NSBHM) \cite{Dutta2015}:

\begin{equation}
\label{equ:NBHM}
 \begin{split}
 \mathcal{\hat{H}}_{NS} &= \mathcal{\hat{H}}_{BH} -T(t) \sum_i \left[\hat{b}_i^{\dagger}\left(\hat{n}_i+\hat{n}_{i+1}\right)\hat{b}_{i+1}+H.c.\right]  \\
 & + \frac{P(t)}{2} \sum_i \left(\hat{b}_i^{\dagger 2} \hat{b}_{i+1}^2 +H.c.\right) +\frac{U_2(t)}{2} \sum_i \hat{n}_i \hat{n}_{i+1}\\
 & -J_2 \sum_i \left(\hat{b}_i^{\dagger} \hat{b}_{i+2} +H.c.\right),
\end{split}
\end{equation}
where $\mathcal{\hat{H}}_{BH}=-J\sum_i (\hat{b}_i^{\dagger} \hat{b}_{i+1}+H.c.) +U(t)/2 \sum_i \hat{n}_i (\hat{n}_i-1) + W\sum_i h_i \hat{n}_i$ is the standard Bose-Hubbard model. $W$ is the amplitude of the disorder potential and the random value $h_i$ at the $i$-th site is obtained from the quasi-periodic lattice (Appendices).
In the NSBHM, $T$ is the density-induced tunnelling, $P$ is the pair tunnelling, $U_2$ is the nearest-neighbour interaction, and $J_2$ denotes the next-nearest-neighbour tunnelling. The modulation of the $y$ lattice induces driving terms to those interaction-relevant terms, including $U$, $T$, $P$ and $U_2$. 
In the ladder systems the driving is realized through the $x$ lattice instead of the $y$ lattice (Appendix A.4). Apart from that, we also include all the additional non-standard terms along the rung direction of the ladder in the simulation.

\begin{figure}[tbp] 
	\centering
	\includegraphics[width=0.49\textwidth]{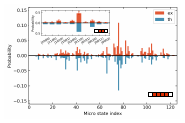}
	\caption{Classical fidelities. For the (6,4) chain, the dimension of Hilbert space is 126. We collected 479 samples to reconstruct the probability distribution and compared it with the numerical simulation, yielding a classical fidelity of $F_c=$ 0.90(2) and a total variance distance (TVD) of $d=$ 0.28(2). The distribution inferred from the experiments samples is plotted in the upper panel and the one obtained from classical simulation is in the lower panel.Inset shows the classical fidelity of the (4,2) chain. Those error bars indicate standard deviation assuming a Poissonian distribution.}
	\label{fig:Fc_64}
\end{figure}

\begin{figure*}[!tpb]
    \includegraphics[width=\linewidth]{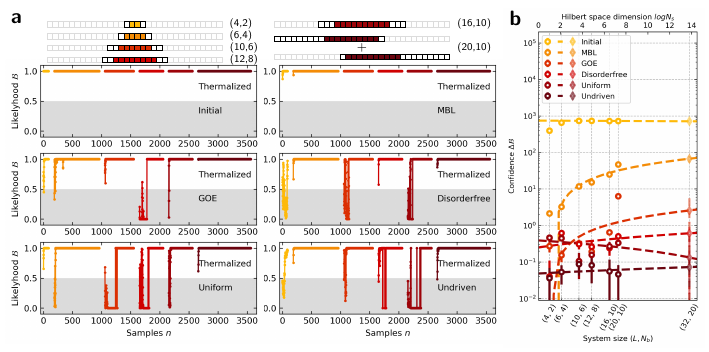}%
    \caption{\label{fig:BayandCon}
    Bayesian tests. 
    (a) Bayesian tests against 6 mock-ups, including initial, MBL, Gaussian orthogonal ensemble (GOE), Disorder-free, Uniform, and Undriven (lower panels, left to right). In each panel, we plot the curve of $\mathcal{B}(\mathrm{ideal},\mathrm{mk})$. The black grids indicate the chain with a given length of $L$ which is cropped from the 32-site chain (grey grids), where bosons are initialized in the middle (filled colored sites). And for the 20-site chain with 10 bosons (20,10) we have two configurations. One is cropped the left and the other from the right boundary of the 32-site chain so that the combination of the two chains cover the whole 32 sites. 
    For all the sampling sets, $ \mathcal{B}(\mathrm{ideal},\mathrm{mk}) >0.5$ and approaches unity so the experimental samples passed all the tests, indicating that the samples are more likely generated by the target thermalized phase. To clearly show the curves, we displace them in the horizontal axis. 
    (b) The confidences of the Bayesian tests. All the confidences are positive (open circles) in the verifiable regime below 20 sites. Error bars are the standard error of the mean (SEM) without error propagation from the calibrated Hubbard parameters and are smaller than the circles if they are being hidden. We linearly extrapolate the confidence (solid lines) to the largest size of 1D $L=32$, $N_b=20$ chain (open diamonds). For the undriven case, the extrapolated Bayesian confidence is $\Delta \mathcal{B}$ = 0.072 $\pm$ 0.051 for the (32,20) system.}
\end{figure*}

Depending on the driving frequency and disorder, there exist three phases: driven thermalized, MBL and prethermalized phases (Fig. \ref{fig:Setup}(b)) and in this work we only focus on the former two phases. What we are interested in is the thermalized phase, which supports the demonstration of quantum advantage. We further compared the different behaviors of the entanglement and correlation between the thermalized and MBL phases in experiments. The chosen parameters are $J/h=166(1)$ Hz, $U/h=422(1)$ Hz, $\omega =200$ Hz, and $W/h=200$ Hz, where $h$ is Planck's constant, lying in the driven thermalized phase. 
For the MBL phase, we increased the disorder to around $W/h=4000$ Hz.
The system evolves under the periodic driving for 10 cycles, corresponding to 50 ms. After that, we freeze the dynamics and expand the atoms in the $y$ direction before imaging \cite{Kaufman2016} to eliminate the pairwise loss due to photo-association in fluorescence imaging. For the number-resolved detection of atoms in both legs of the ladder, one additional step is inserted before the expansion. The atoms are first handed over to the long lattice followed by a barrier projected by DMD to avoid crosstalk between atoms in the two legs.
To precisely model the experiments, we developed various method to calibrate the Hubbard parameters and inserted these calibrations during the data acquisition, which is crucial in the following validations for the samples from experiments (see Appendix B, Fig. \ref{fig:Error_Tol}). 

\section{Classical fidelity}
We first assess the performance of our ultracold atom processor using classical fidelity in small-scale systems. 
Classical fidelity is widely used in boson sampling experiments \cite{Spring2013,Crespi2013,Tillmann2013,Carolan2014,Wang2017c,Zhong2018}, which characterizes the overlap between the probability distribution of the experimental samples and the ideal probability distribution. It is defined as $F_c=\sum_i \sqrt{p_i q_i}$, where $p_i$ is the ideal probability distribution and $q_i$ is the one inferred from the statistics of the samples obtained in experiments. 
In addition, the associated total variance distance (TVD) between the probability distributions is defined as $d=1/2 \sum_i|p_i-q_i|$.
For $L=4$ and $N_b=2$, the dimension of the Hilbert space is 10, which allows efficient reconstruction of the probability distribution in experiments. 
The extracted values are $F_c=0.981\pm0.053$ and $d=0.153\pm0.053$ (Fig. \ref{fig:Fc_64} inset), indicating that the preparation and evolution in the experiments are well captured by the model Eq. \ref{equ:NBHM}. 
Moreover, in a bigger system of 6 sites with 4 atoms we obtained a classical fidelity of 0.90(2) and a TVD of 0.28(2) (Fig. \ref{fig:Fc_64}).
These values are comparable with those in boson sampling experiments \cite{Spring2013,Crespi2013,Tillmann2013,Carolan2014,Wang2017c,Zhong2018}. Please note that the finite sampling effect reduces $F_c$ even for samples generated from the ideal probability distribution \cite{Spring2013}. Thus, the fidelity reported here is underestimated.\\

\section{Bayesian test}
Next, we explore the larger-scale systems where the samples are sparse compared to the Hilbert space. To validate samples generated by the quantum machine, we perform Bayesian hypothesis tests \cite{Bentivegna2014}
against a series of mock-ups. 
This Bayesian method has been used to validate the performance of large-scale boson sampling experiments \cite{Zhong2021a,Madsen2022}.
For an ideal probability distribution $P_{\mathrm{ideal}}$ and a mock-up one $P_{\mathrm{mk}}$, we test the likelihood of the samples from the two probability distributions. Specifically, we define a ratio of the Bayesian likelihood $ \mathcal{B}(\mathrm{ideal},\mathrm{mk})$ as:
\begin{equation}
\label{equ:Bay}
    \mathcal{B}(\mathrm{ideal},\mathrm{mk})_{n}=\frac{\prod_{i=1}^{n} P_{\mathrm{ideal}}(s_i)}{\prod_{i=1}^{n} P_{\mathrm{ideal}}(s_i)+\prod_{i=1}^{n} P_{\mathrm{mk}}(s_i)},
\end{equation}
\noindent where $n$ is the number of the samples and $s_i$ is an instance of the samples. A likelihood ratio $ \mathcal{B}(\mathrm{ideal},\mathrm{mk}) >0.5$ means that the samples are more likely generated from the ideal sampler rather than from the mock-up one.

\begin{figure}[!tbhp]
    \includegraphics[width=0.5\textwidth]{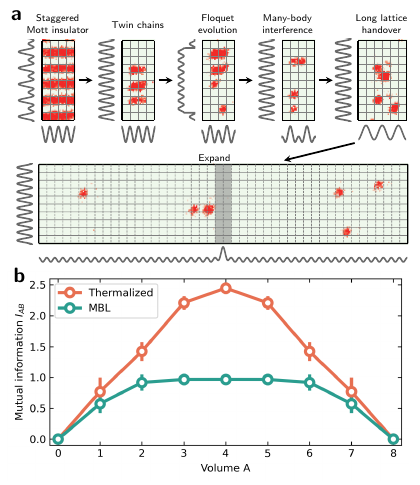}%
    \caption{Entanglement entropy of the driving systems. 
    (a) Schematics of the many-body interferometer. To perform beam-split operations, we prepare two neighbouring copies of chains, each with 3 atoms in 8 sites. After independent evolution, we freeze the dynamics and then apply a double well potential by the $y$ superlattice to perform many-body interference. Finally, the parity is read out to extract the second-order \Renyi entropy after expansion. 
    (b) The entropy of the thermalized phase versus the volume of the subsystem obeys a volume law, whereas it is the area law for the MBL phase. 
    The entanglement data are averaged over all possible combinations of the subsystems with the same system size. Error bars denote the SEM.
    }
    
\label{fig:Ent}
\end{figure}

We present six relevant mock-ups and find that the experimental samples are more likely from the driven thermalized sampler (ideal sampler) rather than all these mock-up samplers for varied sizes of the system (see Fig. \ref{fig:BayandCon}(a)). After dozens of samples, the likelihoods converge to unity, ruling out the mock-ups. To gain further information about the likelihood, we introduce the confidence of the Bayesian test $\Delta \mathcal{B} (\mathrm{ideal},\mathrm{mk})_{n} = \log \left[ \prod_{i=1}^{n} P_{\mathrm{ideal}}(s_i)/\prod_{i=1}^{n} P_{\mathrm{mk}}(s_i)\right]/n$, which is positive if the samples are more likely generated from the ideal sampler. For all the mock-ups, the confidences are positive in the classical verifiable regime of up to $L=20$, $N_b=10$. 
Note that for (20,10) system we have two realizations starting from each end of the chain to cover all the 32 sites. The two displaced patches help accounting for the inhomogeneity of the 32-site system induced by the lattice potentials.
We extrapolate the confidences to the advantage regime of 32 sites (Fig. \ref{fig:BayandCon}(b)) and find that the confidences are positive, inferring that the samples are more likely from the driven thermalized phase. Using Schr\"odinger evolution (SE) algorithm, it would take 2 months to obtain the confidence of the samples in the $L=32$, $N_b=20$ system on \textit{Hanhai20} clusters (28,800 cores, 138 terabytes) at USTC if the memory resources are sufficient. Even in the currently available fastest supercomputer, \textit{Frontier}, the task of obtaining a sample would take more than 2,500 seconds, yielding a quantum speedup of 5-fold compared to the \textit{Frontier} supercomputer. Extending to the two-leg ladder system of up to $L= 64$, $N_b=20$, the quantum speedup will grow by at least 3 orders of magnitude even taking the infidelities of the experimental data into account. Although the quantum advantage demonstrated here is moderate and there are possible improvements of classical algorithms to reduce the cost in the future, we still expect that the quantum-classical gap will be enlarged consistently by upgrades of larger quantum processors. \\

\begin{figure*}
    \includegraphics[width=\textwidth]{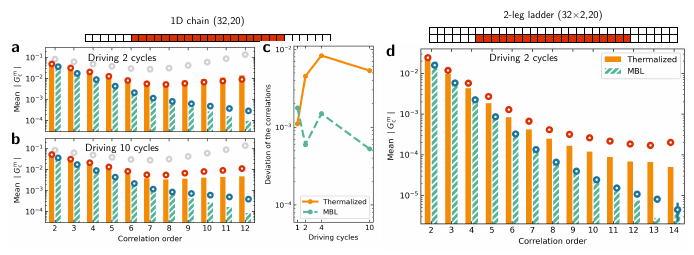}%
    \caption{\label{fig:cor}
    Multi-point density correlations in a Floquet-driven Bose-Hubbard system.
    (a)-(b), We extract multi-point density correlation functions from the experimental samples (open circles) for the 32-site chains. The correlations distinguish the thermalized phase from the MBL phase by enhanced correlation in higher orders. We compare the correlations with 2 cycles (a) and 10 cycles driving (b). Correlations of up to 12th-order are shown. For comparison, the prediction from an infinite-temperature ensemble is shown in gray circles. (c) In this regime, the approximate simulation algorithm of time-dependent variational principle (TDVP) could not reproduce the correlation faithfully for the thermalized phase as we increase the driving cycles. We show the deviations between experimental data and classical simulation results from the TDVP with bond dimension of 200. Error bars are the SEM and are hidden if smaller than the circles. Bars (slashed bars) denote the predicted value of the thermalized phase (MBL phase) from the SE simulations without any free parameters. The average value of $|G_{\mathrm{c}}^m|$ takes over all the possible combinations of the subsystems.
    (d) The correlations in a two-leg ladder of 64 sites and 20 atoms with 2 cycles driving. We could see clear distinctions between the thermalized and MBL phases as well. And now the deviations of the thermalized phase between the experimental data and TDVP simulation are more pronounced in comparison with the 1D case. To remove finite sampling effects, we set the number of samples used in the analysis for each case to exactly the same. For the 1D system in (a), (b) and (c), it is 65 and 100 for the ladder system in (d).}
    
\end{figure*}

\section{Entanglement entropy}
A key feature of chaotic dynamics is the fast scrambling of entanglement. 
We perform many-body interference to measure the second-order \Renyi entanglement entropy \cite{Islam2015,Kaufman2016,Brydges2019,Bluvstein2022,Tajik2023,Joshi2023,Bluvstein2024} $S_2^A=-\ln \mathrm{Tr}(\rho_A^2)$, where $\rho_A $ is the reduced density matrix of subsystem $A$. Two copies of the chain are selected by the addressing beam. After driving for a varied number of cycles, we freeze the dynamics by ramping up the $x$ short lattice to 51$E_{rS}$. 
Then the $y$ long lattice is ramped up, implementing a balanced double well together with the lowered $y$ short lattice for the many-body interference (Fig. \ref{fig:Ent}(a)).
After finishing the  beam splitter operation, we freeze the dynamics in the $y$ direction and image the atoms with expanding in the two separated areas similar in the aforementioned ladder sampling experiments. 
The parity of the output in two ports could be obtained microscopically, from which we extract the purity and entanglement entropy on various sizes of the subsystems.

Fig. \ref{fig:Ent}(b) shows the size dependence of the entanglement entropy measured by two copies of the $L=8$, $N_b=3$ chains. The mutual information defined as $I_{AB}=S_2^A+S_2^B-S_2^{AB}$ is used widely in condensed matter physics and quantum information science. In the thermalized phase, we observe a volume law of the mutual information, suggesting that there is no efficient MPS representation for highly entangled states as the size of the system grows \cite{Schuch2008}. In contrast, it obeys an area law in the MBL phase, which is tractable with MPS for larger system sizes on classical computers. \\

\section{Multi-point density correlation functions}
Based on the sampling results, we further analyse the multi-point correlation functions \cite{Hodgman2011,Dall2013,Schweigler2017,Rispoli2019a,Koepsell2020b,Zhong2021a} of the many-body system. In a Floquet system, multi-particle interactions emerge, which marks the breakdown of the Magnus expansion \cite{Thanasilp2020,Magnus1954,Ponte2015} and leads to enhanced high-order correlations. 
We exploit the multi-point density correlations to probe the many-body nature of the system. 
For the two-point case, the connected correlation is defined as  $G_{\mathrm{c}}^2(x_1,x_2)=G_{\mathrm{tot}}^2(x_1,x_2)-G_{\mathrm{dis}}^2(x_1,x_2)=\braket{\hat{n}_1 \hat{n}_2}-\braket{\hat{n}_1}\braket{\hat{n}_2}$, where $\braket{\hat{n}_1}$ and $\braket{\hat{n}_2}$ are the density of site $x_1$ and $x_2$ respectively, and $\braket{\hat{n}_1 \hat{n}_2}$ is the total correlation between the two sites, including both the connected and disconnected (lower-order) parts.
Higher-order $m$-point multi-correlation functions of density are defined recursively as: 

\begin{equation}
\label{equ:Cor}
    G_{\mathrm{c}}^m(\mathbf{x})=G_{\mathrm{tot}}^m(\mathbf{x})-G_{\mathrm{dis}}^m(\mathbf{x}),
\end{equation}
where $G_{\mathrm{c}}^m(\mathbf{x})$ is the connected correlation function applying to sites $\mathbf{x}$, and $G_{\mathrm{dis}}^m(\mathbf{x})$ denotes the contributions of all the lower-order correlations (Appendices). 

From the atom-number-resolved readout, we extract multi-point correlations up to 14th-order in the largest size of $L=64$. The connected correlations are genuine many-body correlations, signalling the multipartite entanglement \cite{Rispoli2019a}. As shown in Fig. \ref{fig:cor}, the enhanced higher-order correlations in the thermalized phase contrast with those in the MBL phase. The experimental data are consistent with predictions from simulations of the SE in systems up to 20 sites. Beyond that, it is infeasible for the SE algorithm. Therefore, we employ instead the MPS-based time-dependent variational principle (TDVP) algorithm for the classical simulation in the 32-site chain and $32\times2$ ladder as approximations of the experiments.
To account for the finite sampling effect, we set the number of samples to be exactly the same for a fair comparison between experiments and classical simulations and between thermalized phase and MBL phase.

Firstly, we find significant distinctions between the thermalized phase and MBL phase in terms of the multi-point correlations. Those correlations are enhanced in the thermalized phase compared to that of the MBL phase especially in the higher orders. 
Second, the predicted values from TDVP simulations are deviating more and more away from the experimental data
as we increase the driving cycles in the thermalized phase (See Fig. \ref{fig:cor}(c) and Fig. \ref{fig:Corcycle}). 
By contrast, in the MBL phase the TDVP method is always able to capture the physics faithfully even after longer time dynamics. 
Lastly, we also measured the intensity noises and phase noises of the lattices in our previous works \cite{Li2021,Wang2022} and those noises are negligible to exciting atoms and comparable to previous reported numbers \cite{Blatt2015}. This can be further confirmed by checking the correlations over various time when switching off the driving. There is no obvious deviation over time in terms of the multi-point correlations, which suggests negligible heating injected to the system (See Fig. \ref{fig:cor_undriven}).

Thereby, we attribute the deviations mainly to the underestimation of those correlations by the TDVP simulations rather than the increase induced by heating during the longer driving in the experiments. Because the heating induced by the lasers would be much larger in the MBL phase which has a much stronger disorder light field. 
This underestimation is due to the cut-off of the bond dimension in the TDVP simulation which involves less entanglement and cannot capture the higher-order correlations (Appendices). This argument could also be further confirmed by the observation that the deviation is size-dependent. On one hand, the experimental data for the thermalized phase agrees well with the predictions by exact method of SE in the smaller size systems for 10 cycles of driving. 
For (32,20) system, a bond dimension of 200 for TDVP is still capable to reproduce comparable results for experiments within 2 driving cycles. As we increased the driving cycles, the errors accumulated due to more and more truncations involved in the TDVP simulation. Consequently, for 10 driving cycles it shown a decrease systematically in the higher order correlations which is clearly a signal of the breakdown for the approximation simulation with insufficient bond dimension.
However, when we go to the largest system of $L=64$, $N_b=20$, the predictions from TDVP are much lower than the experimental data even with 2 cycles driving, while it is consistent with the 32-site chain experiments for the same driving cycle number.
It indicates that for a larger system increasing the bond dimension significantly is required for the classical simulation to get comparable results from the quantum machine.


Moreover, higher-order correlations are impossibly accessed by samples generated by mock-up samplers merely containing lower-order correlations \cite{Zhong2021a}. This evidence further verifies the hardness for classically simulating this many-body system.\\

\section{Conclusion and outlook}
In summary, we experimentally performed large scale sampling in the driven Bose-Hubbard systems on a quantum processor of ultracold atoms.
We achieved the utilizable quantum speedup over its classical counterpart in simulating the long time dynamics of an interacting many-body quantum system. 
This work establishes the utility of optical lattice quantum simulators for quantitative investigations on many-body dynamics beyond the capability of classical computation, opening the door for practical quantum computational advantage with noisy intermediate-scale quantum devices \cite{Flannigan2022,Daley2022,Trivedi2022}.
For future experiments, one can scale up the size of the problem straightforwardly by generalising to a two-dimensional (2D) system or introducing two types of bosons with different internal states.
The main bottleneck for scaling up in 1D or ladder system is the inhomogeneity of the optical potentials.
The challenge for 2D system is we would suffer from the parity projection detection as now we have no space to expand the atoms in plane. Thus it also raises an open question that whether the complexity of the parity projected sampling degrades or not.

On the other hand, the cycle time could be further reduced by utilizing direct laser cooling the atoms to degenerate \cite{Stellmer2013,Hu2017} or near-degenerate \cite{Phelps2020}.
The techniques demonstrated here could be adapted to implement quantum-enhanced measurements  \cite{Huang2022} with ultracold atoms.
The optimal experimental parameters could be found via Hamiltonian learning \cite{Wang2017,Bairey2019,Li2020,Ott2024} in a much larger dimension of the Hilbert space.
Based on the programmabilities of the disorder and driving, we could explore the emergent geometries of the system \cite{Periwal2021} from the multi-point correlations in the quantum machine.
In addition, the celebrated level statistics of the thermalized and MBL phases could also be probed via many-body spectroscopy \cite{Roushan2017}.\\

\smallskip
\begin{acknowledgements}
We thank Pan Zhang, Honghui Shang, Xiao-Yu Dong, Soonwon Choi and Daniel Mark for helpful discussions. This work was supported by the National Natural Science Foundation of China (Grant No. 12125409), the  Innovation Program for Quantum Science and Technology (2021ZD0302000), the Anhui Initiative in Quantum Information Technologies. We thank the USTC supercomputing center for providing computational resources for this project.
YGZ acknowledged the support by the China Postdoctoral Science Foundation (2023TQ0102), the Fundamental Research Funds for the Central Universities, and the CPS-Huawei MindSpore Fellowship. DA and SP acknowledged support from the National Research Foundation, Singapore, and A*STAR under its CQT Bridging Grant, and the EU HORIZON — Project 101080085 – QCFD.  
\end{acknowledgements}
% ...

\appendix
\section{Experimental sequences and techniques}

\begin{figure*}[htbp] 
	\centering
    \includegraphics[width=1.\textwidth]{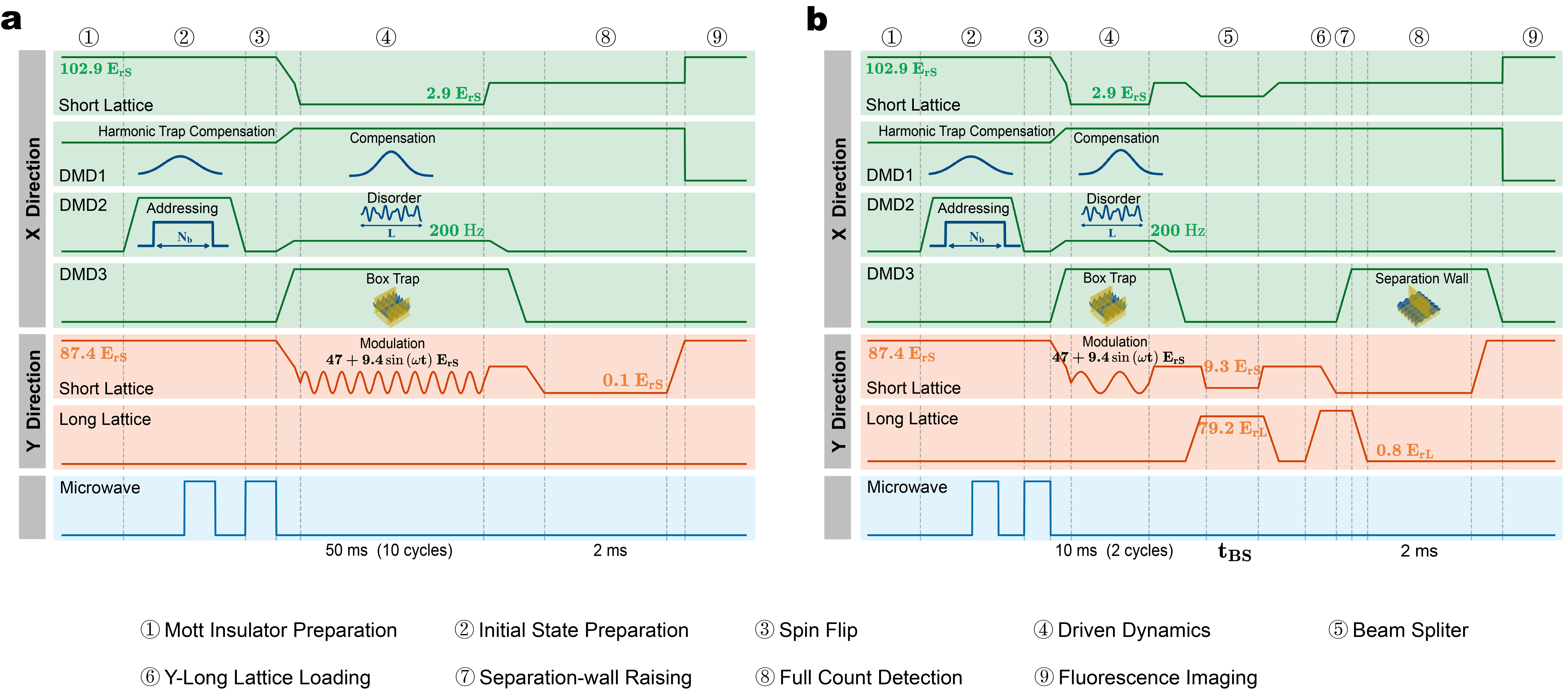}
	\caption{Experimental sequences. (a) Sequence for sampling in the driven thermalized phase of a Hubbard chain. (b) Sequence for probing the entanglement entropy of the driven Hubbard chain.}
	\label{seq_t}
\end{figure*}

\subsection{Sequence for sampling experiments} \label{Seq for 1D}
Our experiments begin with a two-dimensional Bose-Einstein condensate of $^{87}$Rb atoms in the $5S_{1/2} $ $\ket{F=1,m_F=-1}$ state, which is trapped in a single antinode of the $z$ lattice \cite{Xiao2020,Zheng2022a}. The sequence used to draw samples from the thermalized phase is illustrated in Fig. \ref{seq_t}(a).\\
\textbf{Initial state preparation.} During the transition from superfluid to Mott insulator, we perform staggered cooling to obtain a Mott insulator with nearly unity filling \cite{Yang2020}. The staggered potential is applied by the $y$ long lattice, whose trap minima overlap with the minima of the $y$ short lattice. To compensate for the overall envelope induced by red-detuned lattices, we project a blue-detuned anti-trapping potential with DMD1.
The average filling is approximately 0.75 so that the filling is unity in the Mott chains and 0.5 in the reservoir chains, respectively. After the phase transition, we further ramp up the $xy$ short lattices and the $z$ lattice to above 90$E_{rS}$. Then we shift the minima of the $y$ long lattice to overlap with the maxima of the $y$ short lattice and turn on the spin-dependent effect by rotating the polarizations of the lattice beams to introduce a differential light shift between $\ket{F=1,m_F=-1}$ and $\ket{F=2,m_F=-2}$ \cite{Yang2017}. The splitting of the resonant frequency between the Mott chains and reservoir chains is 12 kHz. After selectively flipping the atoms in the reservoir chains, we push out the atoms with a resonant laser of the cycling transition and only the atoms in the Mott chains are retained.
After that, we impose an addressing beam with a magical wavelength ($\lambda = 787.55$ nm) projected by DMD2 to shift the resonant frequency by $12$ kHz between the $\ket{F=1,m_F=-1}$ and $\ket{F=2,m_F=-2}$ states of the addressed atoms \cite{Zheng2022}. We flip the bare atoms to $\ket{F=2,m_F=-2}$ and push out them, akin to cleaning atoms in the reservoir chains.
As a result, we deterministically prepare the initial $1\times N_b$ state in a single chain. Then the retained atoms are flipped to the $\ket{F=2,m_F=-2}$ state for the following experiments.\\
\textbf{Driven dynamics evolution.} 

We then initialize the driving process by suddenly quenching the $x$ lattice to 2.9$E_{rS}$ and modulating the $y$ lattice at the same time. The amplitude of the modulation is 9.4$E_{rS}$. The depth of the $y$ lattice is 47$E_{rS}$, so tunnelling in this direction is suppressed. 
A box trap along the $x$ direction projected by DMD3 confines the atoms in $L$ sites during the evolution. 
In addition, we also remove the inhomogeneities of the chemical potential induced by red-detuned lasers as we did in the superfluid-Mott insulator transition. 
The disorder potential is applied before the evolution.
After ten cycles of driving, we freeze the dynamics by ramping up the $x$ lattice to 51$E_{rS}$.\\
\textbf{Full-count detection.} Due to the light-assisted collisions \cite{Depue1999}, we could only detect the parity of the atomic occupation in fluorescence imaging. To obtain the full count of the final state, we expand the atoms in the $y$ direction by dropping down the $y$ lattice to 0.1$E_{rS}$ \cite{Kaufman2016}. After a time of 2 ms of free tunnelling in the $y$ direction, we pin the atoms to perform fluorescence imaging. Finally, we obtain the samples of the state in an atom-number-resolved way.

\subsection{Sequence for entanglement entropy experiments} \label{Seq for Ent}
To measure the second-order \Renyi entropy, we need to perform many-body interference on two copies. The twins should be prepared in adjacent chains, i.e., in a single double-well formed by the $y$ superlattice. Therefore, we perform the staggered cooling in the $x$ direction instead of the $y$ direction which will result in an alternating structure of the atoms in the $y$ direction.
The sequence to measure the entanglement entropy is illustrated in Fig. \ref{seq_t}(b).\\
\textbf{Initial state preparation.}  
After staggered cooling and pushout of the reservoir chains, we impose an addressing beam to select a $2\times8$ plaquette region. As a result, we create two adjacent chains of state $\ket{00101010}$ in the $x$ direction.\\
\textbf{Beam splitter operation.}  After finishing the driving, we ramp up the $x$ lattice to freeze the dynamics. Then the $y$ long lattice is ramped up to 79.2$E_{rL}$ and the $y$ short lattice is ramped down to 9.3$E_{rS}$, forming a balanced double well. To further reduce the on-site interaction, we lower the depths of the $x$ short lattice and the $z$ lattice to 26$E_{rS}$ and 23$E_{rS}$, respectively. 
The calculated tunnelling rate $J_{BS}/h$ is 967 Hz, and the residue interaction $U_{BS}/h$ is 422 Hz, yielding a contrast of 99\% of the beam splitter operation. After performing a 50$:$50 beam splitter operation on the atoms \cite{Daley2012}, we freeze the atoms by ramping up the $x$ short lattice.\\
\textbf{Full-count detection.} To obtain the occupations of the left and right sites in each double-well, we apply a separation-wall along the $x$ direction in the middle by DMD3. 
Besides, in order to suppressing tunneling in the double-well during detection, we load the atoms into the $y$ long lattice before raising the wall. 
As a results, after free expansion in the $y$ direction,  the atoms in the left sites are spread out to the 50 left sites,  and so are the atoms in the right sites. 
Finally we pin the atoms to perform fluorescence imaging.
The probability that an atom mistakenly crosses the barrier to the other side is $0.7\%$.

\subsection{Measurement of entanglement entropy via many-body interference}
\begin{figure}[tbp] 
	\centering
	\includegraphics[width=0.47\textwidth]{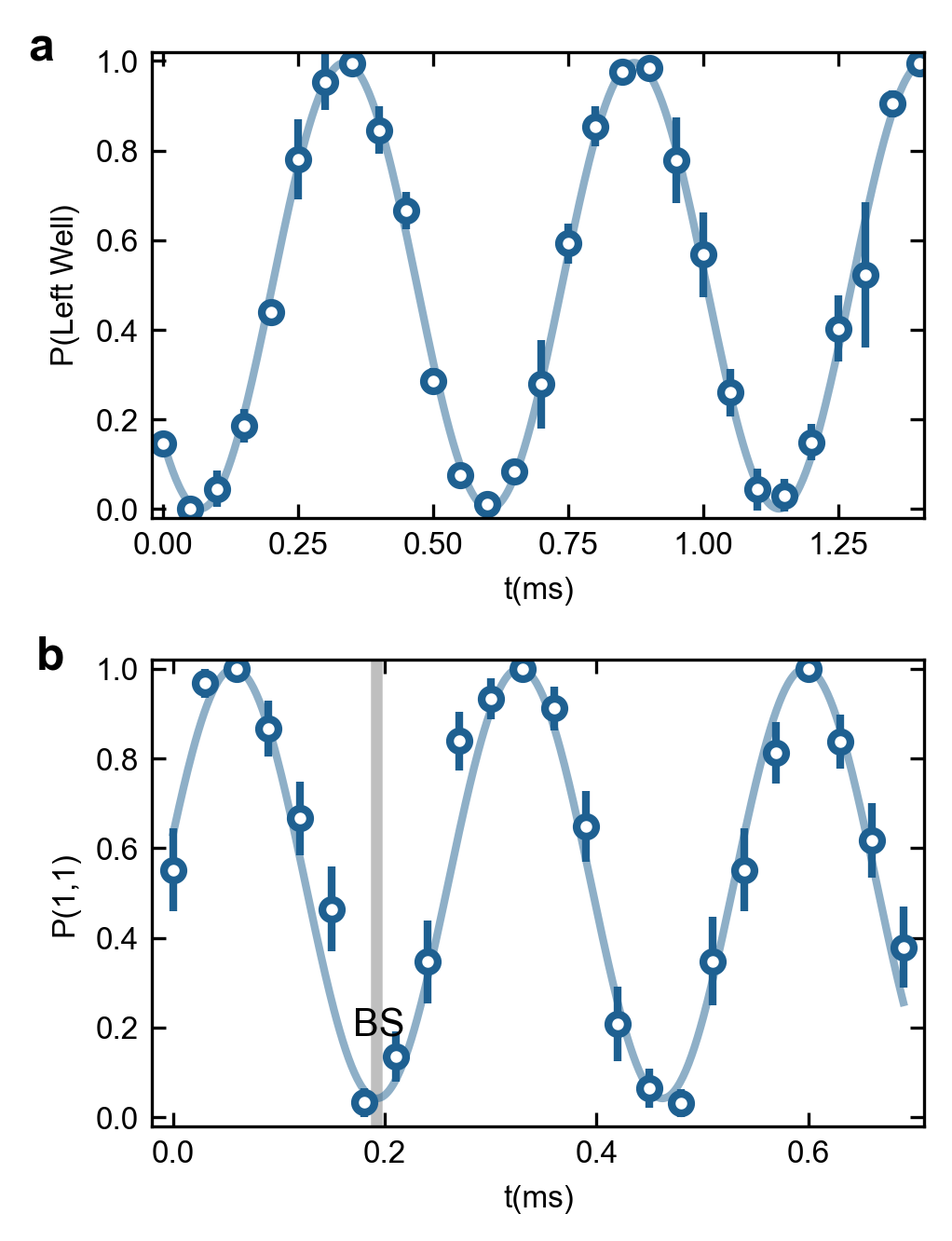}
	\caption{Dynamics of atoms in double well. (a) Single-atom tunnelling process. The atom is initially localized in the left well. (b) Two-atom tunnelling process. The two atoms are initially localized in separate wells. The tunnelling processes start during ramping down of the $y$ short lattice. Hence, the probability $P(Left~Well)$ and $P(1,1)$ are not unity at $t=0$. Error bars denote the SEM.}
	\label{DW_dynamics}
\end{figure}

To calibrate the beam splitter operation, we monitor the tunnelling dynamics of the atom in the double well for both single-atom and two-atom cases. For single atom occupation, it oscillates between the wells with a frequency of $2J_{BS}/h$ as shown in Fig. \ref{DW_dynamics}(a). The measured tunnelling rate is $J_{BS}/h=932(3)$ Hz, and the amplitude is 99(2)\%. 
For the two-atom case, we extract the joint probability $P(1,1)$ of finding two atoms in separate wells. The joint probability $P(1,1)$ oscillates at a frequency of $\sqrt{16J_{BS}^2 + U_{BS}^2} / h \approx 4 J_{BS}/h$.
As shown in Fig. \ref{DW_dynamics}(b), the oscillating frequency is 3704(1) Hz and the contrast is 96(3)\%, which agrees with the prediction. At the minima of the oscillation, the two atoms are in a superposition of $\ket{0,2}$ and $\ket{2,0}$, corresponding to a $50:50$ beam splitter operation.
The first beam splitter time is used in the following experiments. 
At this time, the joint probability is $P(1,1)=0.04(2)$. The purity of the initial Fock state is estimated to be $\langle P_i\rangle=1-2\times P(1,1)=$ 92(4)\%, where $i$ denotes the output ports, indicating a high fidelity of the beam splitter operation. The second-order \Renyi entropy is directly related to the purity \cite{Daley2012} 

\begin{equation}
    S_A^2=-\ln \mathrm{Tr}(\rho_A^2)=-\ln \langle P_i(A) \rangle,
\end{equation}
where $\langle P_i(A) \rangle$ is the purity of subsystem $A$.
These operations are not perfect and introduce extensively classical entropy to experimental data \cite{Islam2015,Kaufman2016,Bluvstein2021a}. 

\begin{figure*}[htbp] 
	\centering
	\includegraphics[width=0.9\textwidth]{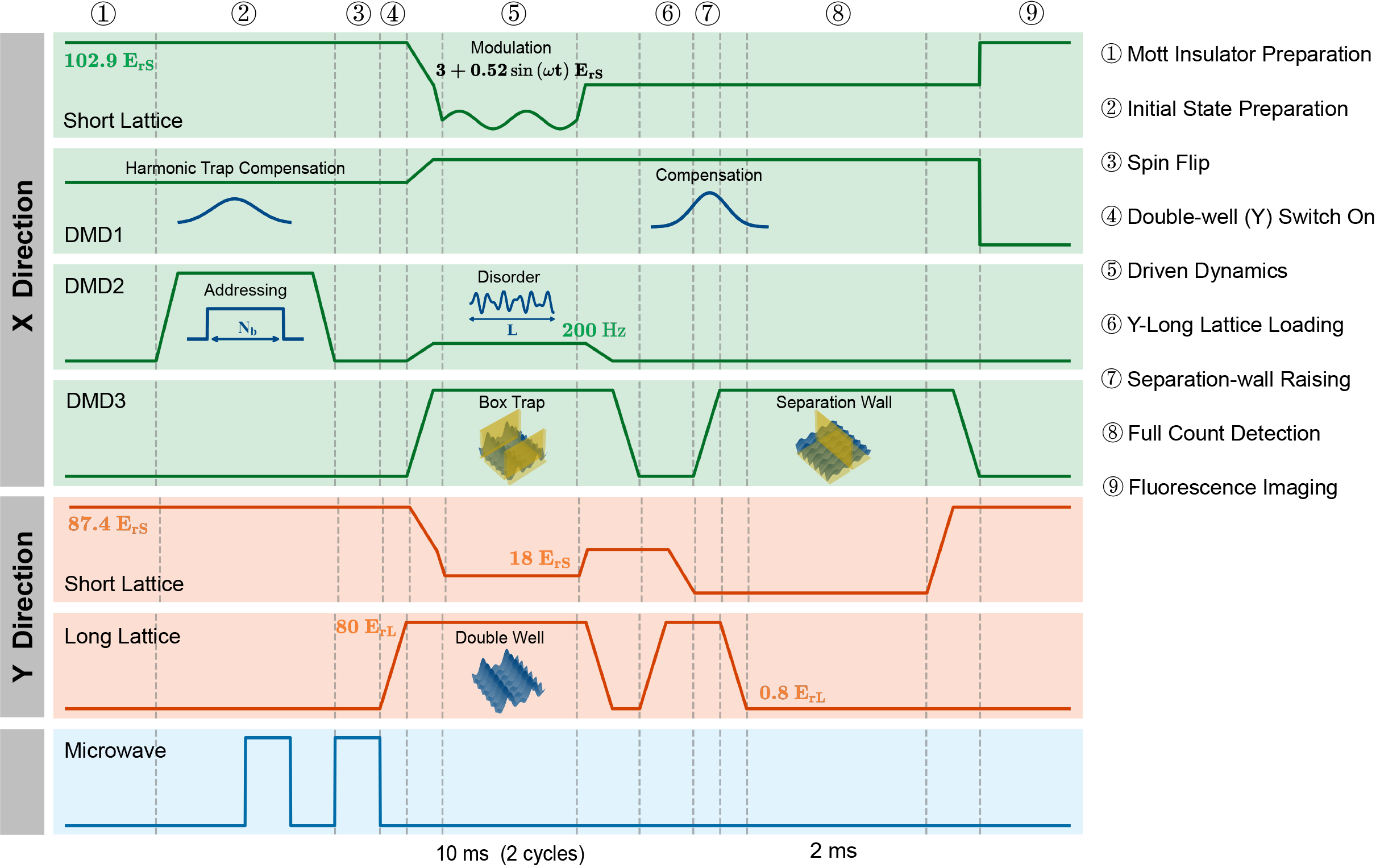}
	\caption{Sequence for sampling in ladder systems. We show some most relevant channels during the sampling in ladder systems.}
	\label{2DLad_Seq}
\end{figure*}

\subsection{Sequence for Sampling in Ladder Systems}
The sequence used to draw samples from the ladder systems is illustrated in Fig. \ref{2DLad_Seq}. The process of initial state preparation is similar to that in Section \ref{Seq for 1D} and the process of full-count detection is similar to that in Section \ref{Seq for Ent}, while the following step is different.\\
\textbf{Driven dynamics evolution of ladder systems.} 
Unlike the 1D sampling process, we ramp up the $y$ long lattice to about 80$E_{rL}$ before suddenly quench the $x$ short lattice to 2.9$E_{rS}$ and the $y$ short lattice to 47$E_{rS}$. Thus, a double-well structure is formed in the $y$ direction while a normal lattice remains in the $x$ direction. 
Here we modulate the $x$ short lattice according to $V_x=2.9\ E_{rS} + 0.52\ E_{rS}\times\sin(\omega t)$ instead of the $y$ short lattice to avoid changing the shape of the double-well. 
A box trap projected by DMD3 confines the atoms in $L$ sites along the $x$ direction so that the target system contains $2\times L$ sites during the evolution.
After two cycles of driving, we freeze the dynamics by ramping up the $x$ short lattice and the $y$ short lattice.

\subsection{Full-count Detection for Ladder Systems}
\begin{figure*}[ht] 
	\centering
	\includegraphics[width=0.95\textwidth]{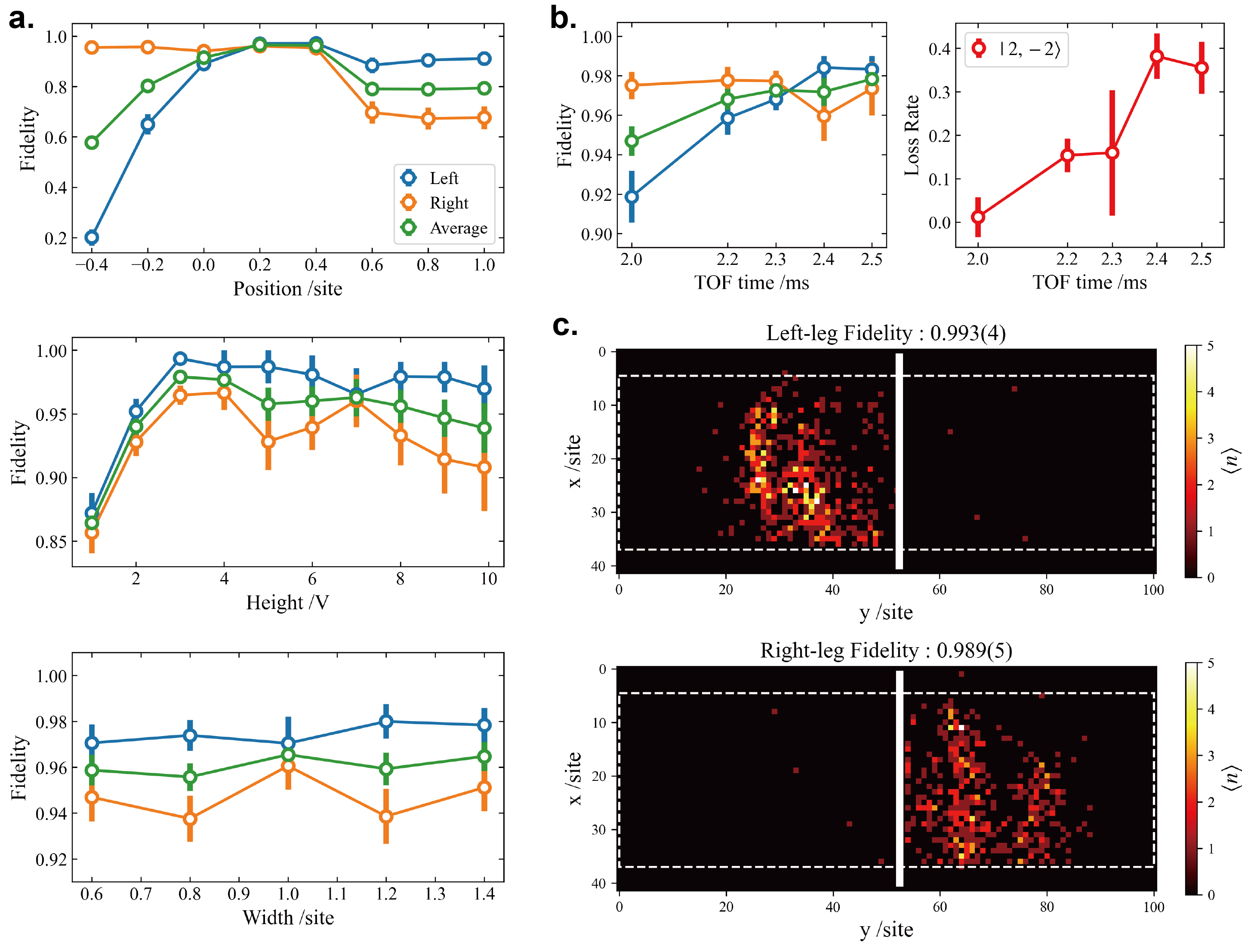}
	\caption{The performances of three different methods for full-count detection. (a) The fidelity to read out the occupations of the left-, right- and average-leg of the first methods. To optimize our first method, we vary the position relative to the superlattice, height and width of the barrier by DMD3. The highest average fidelity is about $98\%$. (b) The fidelity and the loss rate of $\ket{F=2, m_F=-2}$ state in the presence of a gradient magnetic field. To eliminate the mixing between these inward momentum atoms of the two legs, longer expansion time is demanded. However, the atoms in $\ket{F=2, m_F=-2}$ also face an increasing loss rate, which could even reach $40\%$. (c) The performance of the method we finally adopted. The dotted box is the statistical area, and the white solid box indicates the location of the separation-wall. Using this detection method, we are able to achieve an average fidelity of $99.1\%$. Error bars denote the SEM.}
	\label{FCD}
\end{figure*}

To obtain the occupations of the left and right legs in the ladder system, we need to avoid the mixing of the atoms between the two legs during the expansion along the transverse direction. The basic idea is making a barrier between the two legs. To achieve this goal, we tried the following three different methods. Note that in all the three tests, we utilize the spin dependent lattice to address the spin state, serving as a label to distinguish the atoms originally in left or right legs. When we want to know the fidelity of reading out the occupations of the left-leg (right-leg), we will push out the atoms in the right-leg (left-leg) after expanding, so we can get the fidelity from the imaging of the remaining atoms. \\
\textbf{Inserting the separation-wall before expanding.} 
After freezing the dynamics in the short lattice along both directions, we apply a narrow barrier potential along the longitudinal direction between the two legs by DMD3 to separate them. Then we drop down the $y$ short lattice for expanding. To find the optimal performance, we vary the position relative to the superlattice, height and width of the barrier. A thicker barrier is helpful to suppress the tunnelling but the maximal width is limited by the lattice spacing. The results are shown in Fig. \ref{FCD}(a). The highest average fidelity of about $98\%$. \\
\textbf{Expanding in the presence of a gradient magnetic field.} 
After freezing the dynamics, we turn on the spin-dependent effect and flip the atoms on the right leg. Therefore in the double-well, the atoms on left leg is $\ket{F=1, m_F=-1}$ and on the other leg is $\ket{F=2, m_F=-2}$. Afterwards we ramp up a gradient along the transverse direction to guide the movement of the two states during the expansion. These two different states move towards the high field and the low field respectively. Due to the initial uniform momentum distribution, part of the atoms in the two legs will move towards each other. Thus, a stronger gradient or longer expansion time is demanded to eliminate the mixing between these inward momentum atoms of the two legs. However, this will raise another issue that the atoms with outward momentum will be accelerated too much to go outside the view. A trade-off is made to obtain optimal fidelity without too much loss of the atoms (Figure. \ref{FCD}(b)).\\
\textbf{Handover to $y$ long lattice.} 
This approach is an upgrade version of the first method. As we mentioned before the maximal width of the barrier is limited by the lattice spacing. Therefore prior to the ramping up of the barrier potential, we hand over the atoms from the short lattice to the long lattice which has a twice spacing compared to the short one. A larger lattice spacing allows a broader barrier as well as larger tolerance over the drift of the relative position between the barrier and the lattice. Fig. \ref{FCD}(c) shows the statistical results after multiple measurements. The dotted box is the statistical area, and the white solid box indicates the location of the separation-wall. Using this detection method, we are able to achieve an average fidelity of $99.10\%$.

\subsection{Limitations in the experiments}
In this experiment, the decoherence processes include heating from laser intensity noise, scattering of background gas. The intensity noise in the lattice lights could heat the atoms to high bands and the scattering of background gas will induce atom loss. Owing to the near resonant light for the disorder potential, spontaneous scattering process also contributes the decoherence during the evolution especially in the MBL phase where the disorder potential consumes more power of the light.
In the experiments of thermalized phases, the longest driving time is 50 ms which is much shorter than the lifetime bounded by the aforementioned decoherence mechanisms. Instead, it is mainly limited by the coherent deviations of the Hubbard parameters during the evolution. The evolution remains coherent but the deviation of those parameters results a different final state and thereby breaks down the Bayesian tests which relies on precise knowledge of the probability distribution. 
So, we have to calibrate the parameters frequently and lock the phase of disorder potential to the lattices during the sampling experiments. In the experiments of MBL phases, as the disorder potential is much larger the spontaneous scattering heats the atoms. Those atoms are mobile even in the presence of a deep disorder potential and accounts for the increase in the multi-point correlations of MBL compared with numerical simulations as shown in Fig. \ref{fig:cor}. In the ladder experiments, another deviation rises which comes from the drifting relative phase between the long lattice and short lattice to form the double-well potentials along the transverse direction.

\section{Calibrations of Hubbard parameters}
\subsection{Tunnelling and on-site interaction}
\begin{figure}[tbp] 
	\centering
	\includegraphics[width=0.47\textwidth]{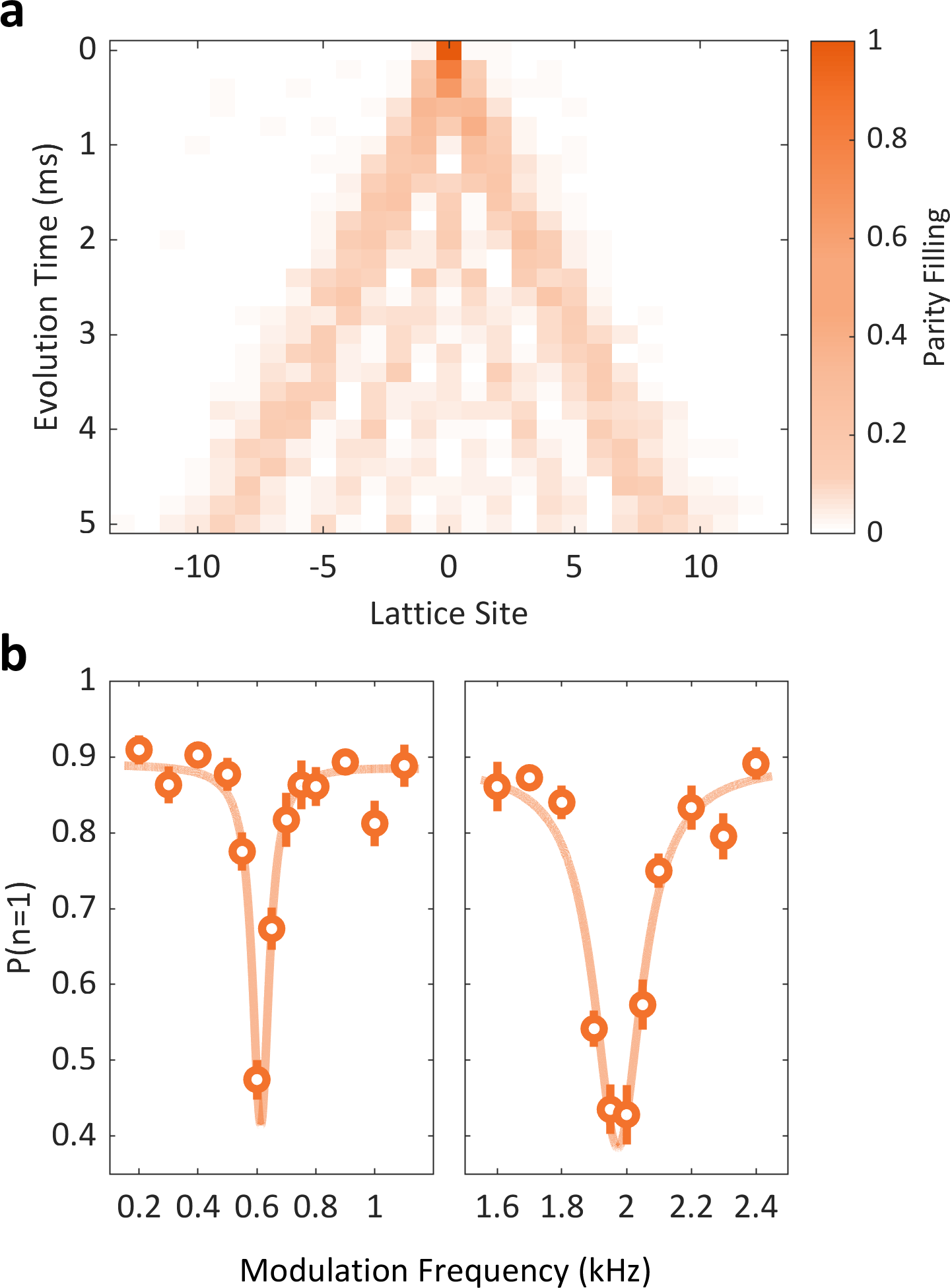}
	\caption{Calibrations of Hubbard parameters $ J $ and $ U $. (a) Quantum walk of a single atom. The result shows $ J/h = 166(1) $ Hz. (b) Plot of the population fraction versus the modulation frequency. The result shows two expected dips at $ E_{tilt} \pm U $ with $ U/h = 680(4) $ Hz. Error bars denote the SEM.}
	\label{JUCali}
\end{figure}

We calibrate the tunnelling strength $ J $ and the on-site interaction strength $ U $ using methods similar to those described in Ref. \cite{Lukin2018}. The tunnelling $J$ is measured at a depth of 2.9$E_{rS}$, which is used in the experiments. For the measurement of interaction $U$, it is too shallow that atoms delocalize quickly even when applying a gradient of 30 G/cm in the $x$ direction. Hence, the depth of the $x$ short lattice is set to 10.4$E_{rS}$ during the calibration of $U$. The modulation amplitude is 2.5$E_{rS}$, and the duration of the modulation is 100 ms.
The result is shown in Fig. \ref{JUCali}(b). Two dips correspond to resonances at $ \nu_1 = E_{tilt}+U $ and $ \nu_2 = E_{tilt}-U $, where $E_{tilt}$ is the tilt between two adjacent sites in the $x$ direction induced by the magnetic gradient field.
These results are consistent with the prediction from band calculations. Thus, the other terms included in the NSBHM are derived directly from the band calculations.

\subsection{Dipole potentials imposed by DMDs}
We use three DMDs in the experiments, which are labelled as DMD1, DMD2, and DMD3.
DMD1 is used to compensate for the overall harmonic trapping potential induced by red-detuned lattices.
DMD2 is used for addressing and imposing disorder potential.
We utilize DMD3 to project a box trap for confining atoms during the driving dynamics.

\begin{figure*}[!htbp]
	\centering
	\includegraphics[width=0.95\textwidth]{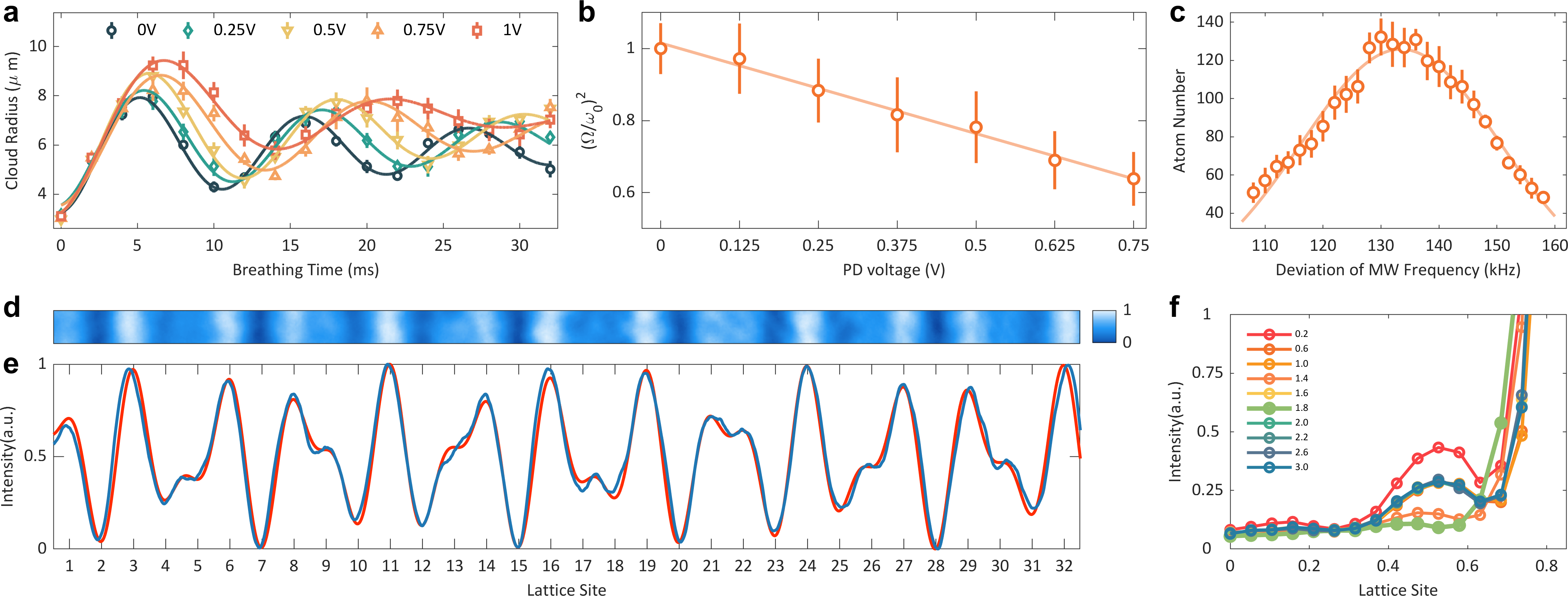}
	\caption{Calibration of potentials projected by DMDs. (a) Breathing-mode oscillations with different powers of the laser projected by DMD1. The solid curves are the fitting results from a damped sinusoidal function. (b) The fitted breathing mode frequency ratios $ (\Omega/\omega_0)^2 $ and the DMD1 power have a linear relation, where $ \omega_0 $ is the breathing mode frequency without the anti-trapping potential. (c) Number of addressed atoms against the induced light shift. The solid curve is a Gaussian fitting, which yields the centre frequency at $ 125.8(3)$ kHz. (d) Image of the disorder potential in $ 1 \times 32 $ sites. (e) Profile of the implemented disorder potential. The measured result (blue curve) is averaged over one site from the above image, which is overlapped with the target one (red curve). The slight deviation between two curves is attributed to the optical aberration and the finite size of the DMD pixels. (f) Profile of the walls in one edge with different sharpness $ \sigma_s $ given by DMD3. The light green line with closed circles represents the optimal sharpness of $ \sigma_s = 1.8 $ pixels on the camera, which minimises the local energy offset to the edge sites. Error bars denote the SEM.}
	\label{DMDAll}
\end{figure*}

\textbf{DMD1.} First, we measure the trapping frequency in the horizontal direction by observing the breathing mode oscillations of the atoms in 2D \cite{Stringari1996}. The frequency of the oscillation corresponds to twice the trapping frequency.
Then, we project an anti-trapping pattern by DMD1 with a 750 nm laser to the atoms. 
The oscillation frequency decreases as we increase the intensity of the laser (see Fig. \ref{DMDAll}(a,b)).
We thus extrapolate the value of the intensity where the trapping frequency is zero, i.e., it is homogeneous in the horizontal direction.

\textbf{DMD2.} The wavelength of the addressing beam is 787.55 nm, and the polarization is circular, which constitutes one magical wavelength between the $D1$ and $D2$ transitions. There is no light shift to $\ket{F=1,m_F=-1}$ yet a red shift to $\ket{F=2,m_F=-2}$. The differential light shift between $\ket{F=1,m_F=-1}$ and $\ket{F=2,m_F=-2}$ could be obtained via microwave (MW) spectroscopy. We project a flattop pattern to the atoms and scan the MW frequency to flip the atoms to $\ket{F=1,m_F=-1}$ before the pushout. The number of retained atoms maximises at the resonant frequency (see Fig. \ref{DMDAll}(c)).

The disorder potential is also imposed by the addressing beam with smaller intensity. The local energy offset is directly related to the extrapolated light shift since the atoms are in state $\ket{F=2,m_F=-2}$.
The disorder potential comes from a quasi-periodic lattice, which is given by

\begin{equation}
\begin{split}
 V_{dis}(x) & = 2W \left\{ (2-\beta)\cos^2\left[\pi(\beta-1)\dfrac{x}{a_S}+\phi\right] \right.  \\
 & \left.+(\beta-1)\cos^2\left[\pi(\beta-2)\dfrac{x}{a_S}+\phi \right] \right\},
\end{split}
\label{equ:dis}
\end{equation}
where $\beta=(\sqrt{5}-1)/2$ is the golden ratio. This type of disorder potential takes the advantage that it is flat in the lattice sites and is immune to the relative position drift between the DMD2 pattern and the lattice phase \cite{Lukin2018}. We choose three instances of disorder potentials with $\phi=$0.1$\pi$, 0.4$\pi$ and 0.7$\pi$.

\textbf{DMD3.} DMD3 projects a box trap potential to confine the atoms during driving. The wavelength of the light is also 750 nm. The pattern of the wall is given by a flattop functions

\begin{equation}
    I = \mathrm{erf}\left(\frac{x-x_0+\sigma_{w}}{\sigma_s}\right)+\mathrm{erf}\left(-\frac{x-x_0-\sigma_{w}}{\sigma_s}\right),
    \label{equ:wal}
\end{equation} 
where $ \mathrm{erf}(\cdots)$ is the Gaussian error function. The sharpness is determined by $ \sigma_s $ and 2$\sigma_w$ is the width of the wall. To minimise the energy offset in the edge sites close to the wall, we optimise the wall by varying the sharpness.
Fig. \ref{DMDAll}(f) shows the profile of the walls in the edge sites with different sharpness parameters. 
With the optimal value, the local energy offset in the edge sites is approximately 150 Hz while the height of the wall is $\sim$10$E_{rS}$. The width of the wall is 6 sites, suppressing atoms from tunnelling outside the walls.

\begin{figure}[tbp] 
	\centering
	\includegraphics[width=0.47\textwidth]{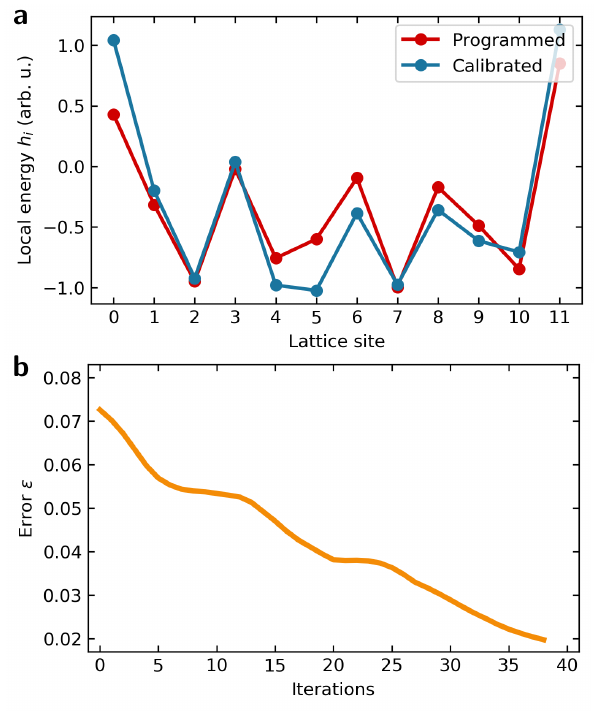} 
	\caption{Calibration of chemical potential. (a) Red circles are programmed disorder potentials for a 12-site chain. Note that the boundary sites are lifted by the wall. Green circles are calibrated local potentials. The deviations between them are attributed to dipole potentials induced by other lasers. (b) Error of density profile decreases versus iterations.}
	\label{fig:Chemi}
\end{figure}

\begin{figure}[tbp] 
	\centering
	\includegraphics[width=0.47\textwidth]{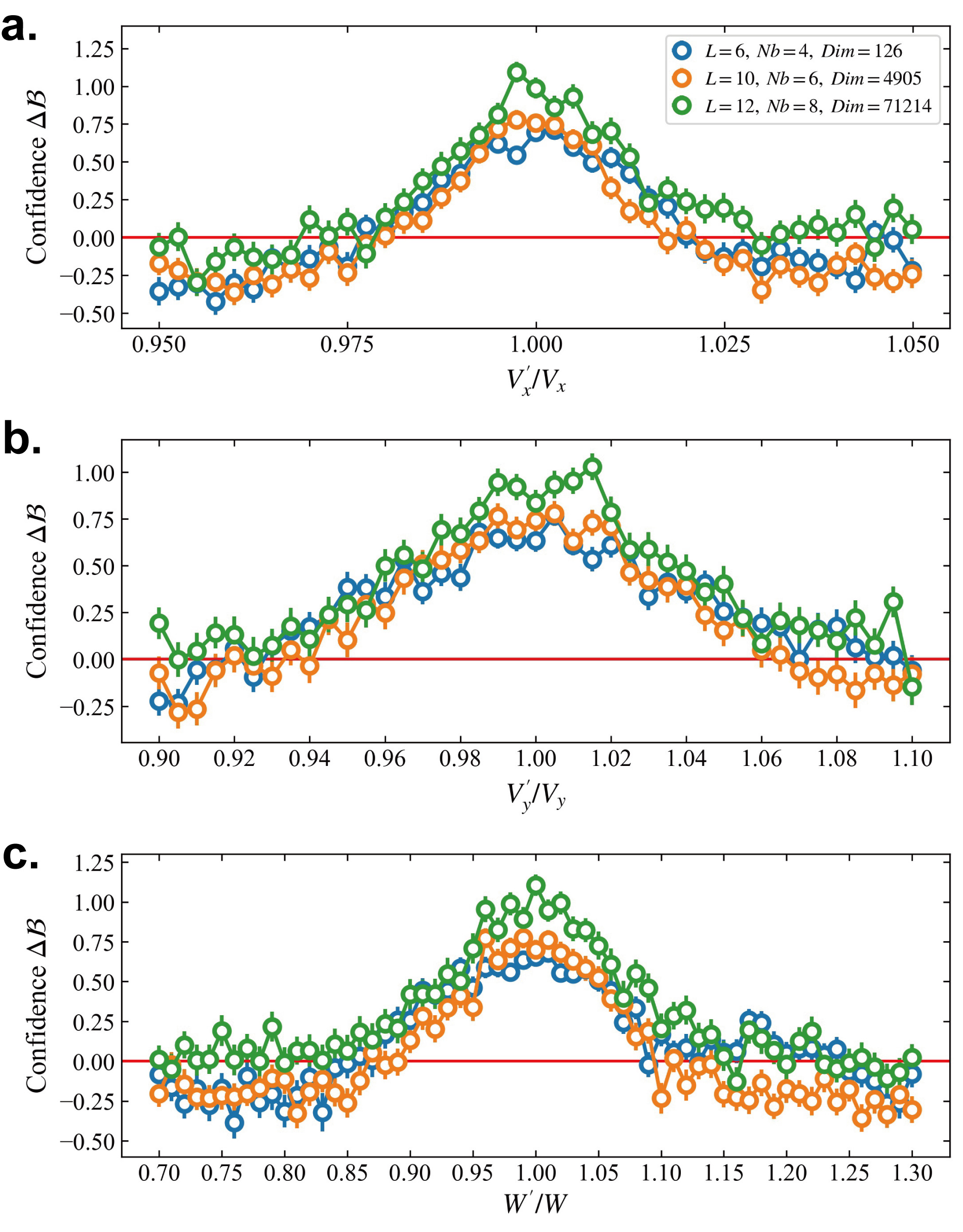}
	\caption{Calibration error tolerance of $V_x,V_y$ and $W$. We simulate sampling process under offset parameters and calculate the Bayesian confidence $\Delta \mathcal{B}$ against uniform distribution for about 300 samples. $\Delta \mathcal{B}>0$ implies the samples can pass the Bayesian test. Error bars denote the SEM.}
	\label{fig:Error_Tol}
\end{figure}

\subsection{Local energy offset}\label{chemit}

Even though the disorder potential could be monitored from the camera, the potential experienced by atoms is the combination of all the lattices and the projected lasers. Hence, we calibrate the potential in the atom plane via an \textit{in situ} method. Thanks to the capability of the site- and number-resolved readout, we could extract the density profile of the chain, which depends on the chemical potential under local density approximation. The local energy shift at each site can be obtained from the self-consistent iterations until the error $\epsilon$ between the density profiles from the experiments and simulations is sufficiently small.
We insert calibration shots during the collection of the samples. The calibration shots share the same parameters with the sampling experiments except without driving. For smaller sizes ($L\le 16$), we exploit SE, and for $L\ge20$, we use TDVP for the iteration, respectively. We have compared the density profiles given by SE and TDVP for smaller sizes and found that the deviations were negligible. In practice, we need dozens of iterations to converge at an error of $\epsilon <$0.02 (see Fig. \ref{fig:Chemi}). 

However, the aforementioned calibration method of the local energy shift breaks down in the MBL phase due to the lack of mobility of the atoms. We could employ the MW spectroscopy in the presence of the strong disorder since the tune-out magic wavelength of the beam would induce differential light shift between $\ket{F=1,m_F=-1}$ and $\ket{F=2,m_F=-2}$ atoms \cite{Choi2016}. While this approach omits the energy shift from other lights, those contributions are negligible compared to the dominant disorder potential projected by DMD2 in the MBL phase.

\subsection{Calibration error tolerance}
The accuracy of the parameter calibration will affect whether the Bayesian test is passed or not. To validate the samples, one must be able to get very precise knowledge of the dynamics happening in the experiment system. As the dynamics is chaotic, any tiny errors in the parameters of the model will give a totally different probability distribution and thereby mess up the validation such as Bayesian tests or cross entropy benchmark. In this subsection, we investigate the error tolerance of Bayesian confidence by numerical simulation. For the ideal probability distribution, the depths of lattices and the strength of disorder are $V_x=3 E_{rS}, V_y=47 E_{rS}, W=200$ Hz respectively. 
First, we get 300 samples directly from the probability distribution $Pr(S)$ without any deviation of the parameters. Then we do Baysian tests against uniform distribution using a deviated probability distribution $Pr^\prime(S)$, which is obtained from classical simulation by intentionally introducing errors to one of the parameters, e.g., $V^\prime_x=(1+\epsilon)V_x$.
In Fig. \ref{fig:Error_Tol}, we plot the test results in terms of the Bayesian confidence $\Delta \mathcal{B}$ for three system sizes and find that the error tolerance interval is independent of system size. Among the above parameters, the one with the smallest tolerance interval is $V_x$. The Bayesian test will fail when the offset $\left|V_x^{\prime}-V_x\right|/V_x$ exceeds $2\%$, imposing an upper bound of the errors in the calibrated experimental parameters.

\subsection{Bayesian tests on samples from MBL phase}
Those aforementioned calibration procedure works well for thermalized phases. As a complementary check, we did the similar calibration of the parameters for MBL phases and tested the various Bayesian hypotheses as well. We tried two small systems of (4,2) and (10,6). The disorder potentials are calibrated from density profiles of the experimental samples. Then we can get the ideal distributions from classical simulations taking these calibrations into account. It turned out all the tests can be passed except the GOE mockup for (4,2). Here we only sampled one set disorder pattern and it gave reasonable results in MBL phases. 

\begin{figure}[tbp] 
	\centering
	\includegraphics[width=0.47\textwidth]{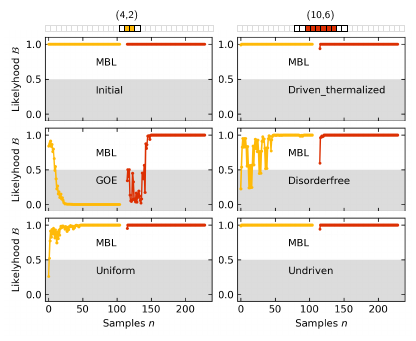}
	\caption{Bayesian tests on samples from MBL phase. We take around one hundred samples for each size. This time MBL distribution is the ideal one while others including thermalized one are mockups with respect to MBL.}
	\label{fig:Bay_Mbl}
\end{figure}

\section{Non-standard Bose-Hubbard model}
\subsection{Non-standard terms}

In our experiment, the lattice depth is as low as $2.9 E_{rS}$ along the $x$ direction. 
For such a shallow optical lattice, the Wannier functions are wider to cover neighbouring sites. The contributions of nearest-neighbour interactions and next-nearest-neighbour processes are not negligible. The NSBHM \cite{Dutta2015} includes these terms in addition to $J$ and $U$, as given in Eq. \ref{equ:NBHM}. Those expressions are listed as follows.
\begin{widetext}
\begin{equation}
\label{equ:NSBH_terms}
\begin{split}
    J=-\int dx w^*_{i}(x) \left[\frac{-\hbar^2 k^2}{2m}\frac{\partial^2}{\partial x^2}+V(x)\right] w_{i+1}(x), \\
    U=g\int dxdydz |w_{i}(x)|^4\times|w_{i}(y)|^4\times|w_{i}(z)|^4, \\
    J_2=-\int dx w^*_{i}(x) \left[-\frac{\hbar^2 k^2}{2m}\frac{\partial^2}{\partial x^2}+V(x)\right] w_{i+2}(x), \\
    U_2=2g\int dxdydz w^*_{i}(x) w^*_{i+1}(x)  w_{i}(x)w_{i+1}(x) \times|w_{i}(y)|^4\times|w_{i}(z)|^4, \\
    T=-g\int dxdydz w^*_{i}(x) w^*_{i}(x)  w_{i}(x)w_{i+1}(x) \times|w_{i}(y)|^4\times|w_{i}(z)|^4, \\
    P=g\int dxdydz w^*_{i}(x) w^*_{i}(x)  w_{i+1}(x)w_{i+1}(x) \times|w_{i}(y)|^4\times|w_{i}(z)|^4,
\end{split}
\end{equation}
\end{widetext}
where $g=4\pi \hbar^2 a/m$ is the contact interaction strength, $a$ is the scattering length and $i$ is the index of the lattice sites along $x$ direction. Note that in our case, the Wannier functions along three axes vary and we only show the terms along $x$ direction as an example.

\begin{figure}[tbp] 
	\centering
	\includegraphics[width=0.47\textwidth]{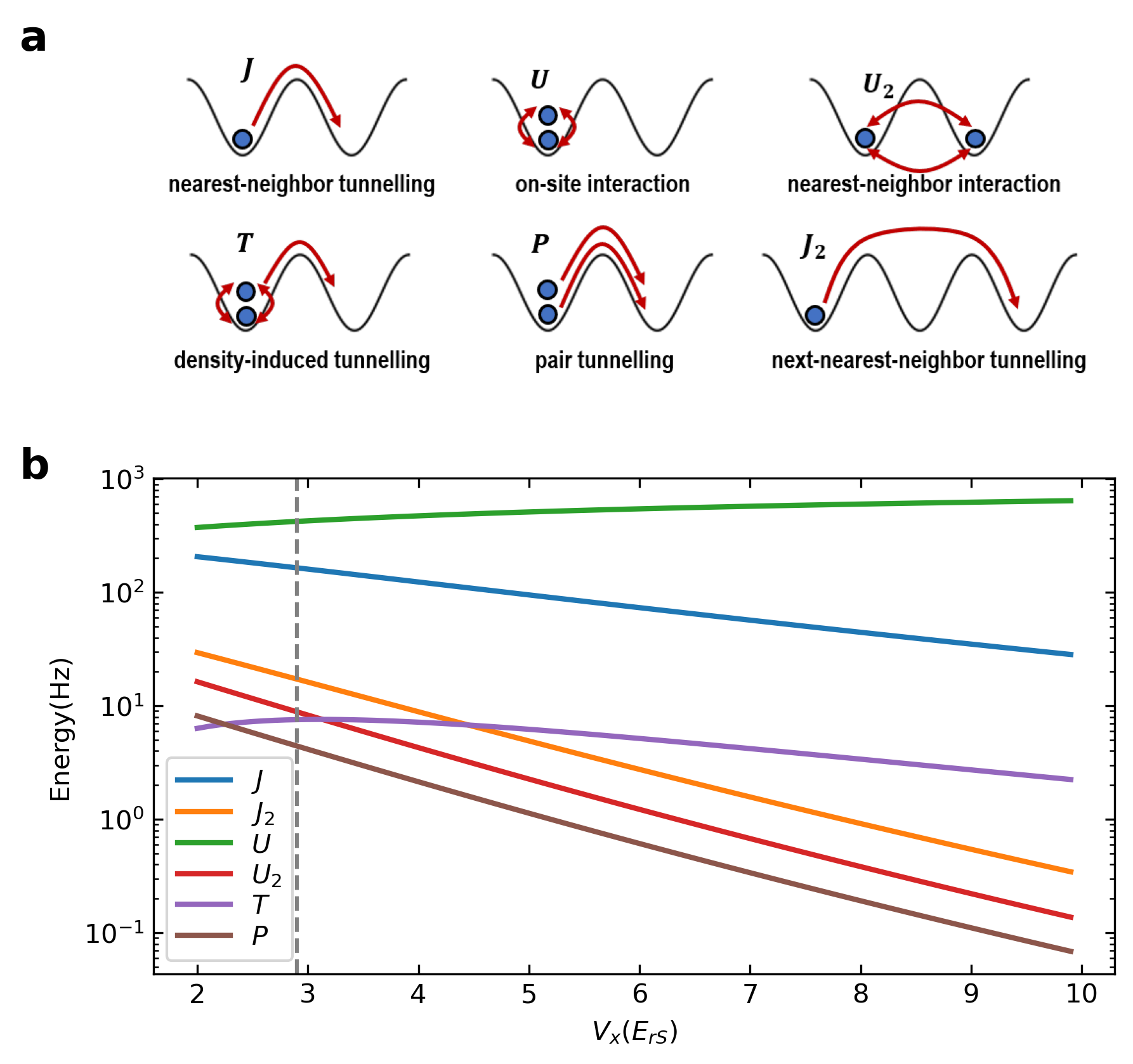}
	\caption{All processes in the non-standard Bose-Hubbard model. (a) $J$ is the nearest-neighbour tunnelling, $U$ is the on-site interaction, $J_2$ is the next-nearest-neighbour tunnelling, $U_2$ is the nearest-neighbour interaction, $T$ is the density-induced tunnelling, and $P$ is the pair tunnelling. (b) The strengths of those processes in the NSBHM versus lattice depth. The additional terms are $\sim $10 percent $J$ at $V_x=2.9 E_{rS}$, while in deep lattices, they are two orders of magnitude smaller than $J$.}
	\label{NSBH}
\end{figure}

Fig. \ref{NSBH}(a) illustrates all contributing processes in this model. Here, $U_2$ is the nearest-neighbour interaction, $T$ is the density-induced tunnelling, $P$ is the pair tunnelling, and $J_2$ denotes the next-nearest-neighbour tunnelling. 

\renewcommand\arraystretch{2}
\begin{table}
	\centering
	\begin{tabular}{cc}
	\hline
	\hline
		Term & Value (Hz) \\
		\hline
		nearest-neighbour tunnelling, $J/h$ & 166 \\
		on-site interaction, $U/h$ & 422 \\
		next-nearest-neighbour tunnelling, $J_2/h$ & 18 \\
		nearest-neighbour interaction, $U_2/h$ & 9 \\	
		density-induced tunnelling, $T/h$ & 8 \\
		pair tunnelling, $P/h$ & 5 \\
		\hline
		\hline
		
	\end{tabular}
	\caption{Parameters in the experiment. These parameters are calculated for $V_x=2.9 E_{rS}$, $V_y=47.0E_{rS}$ and $V_z=45.4E_{rS}$}
	\label{NSBH3}
\end{table}

\begin{figure}[htbp] 
	\centering
	\includegraphics[width=0.47\textwidth]{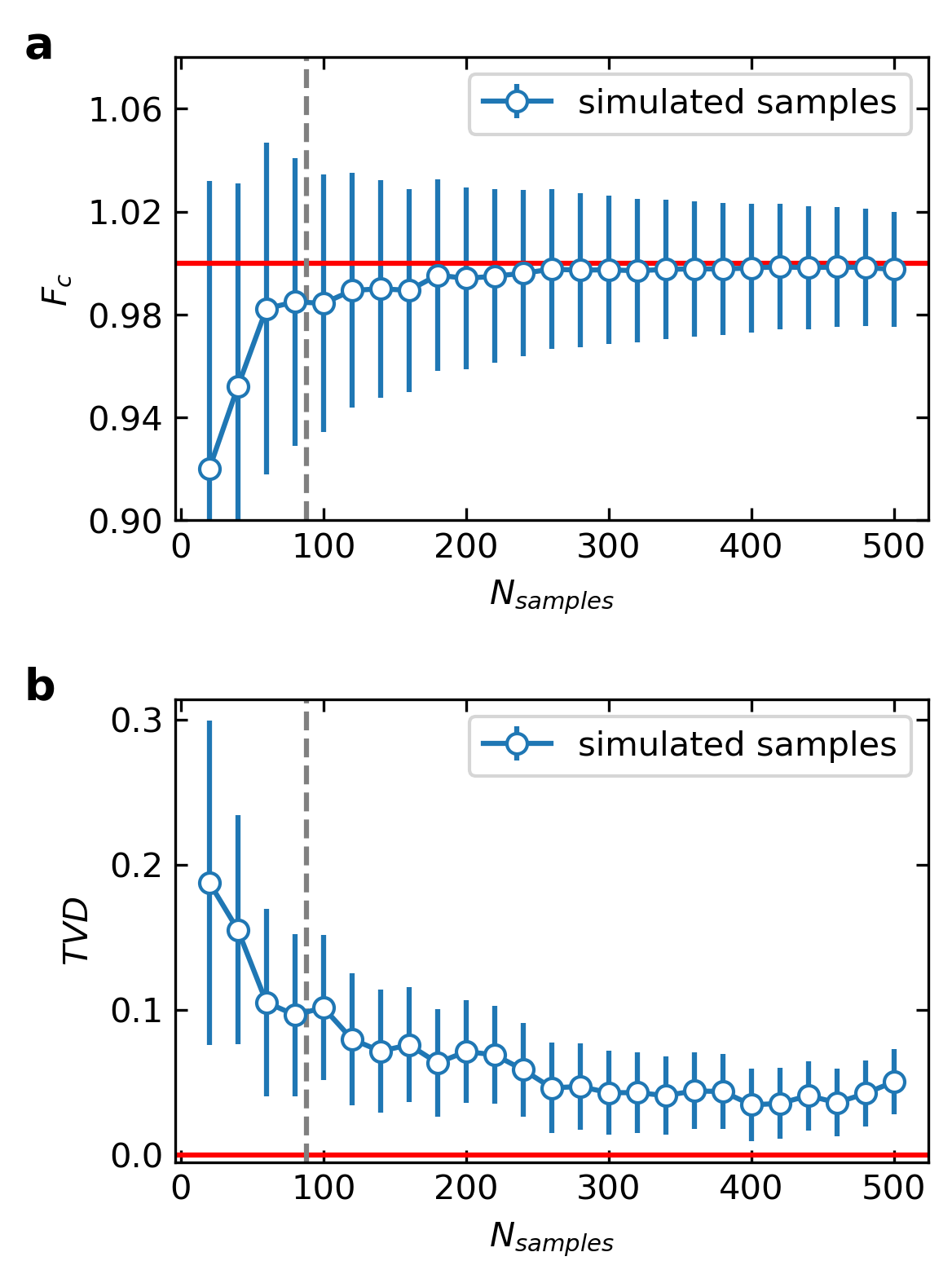}
	\caption{Classical fidelity and TVD for simulated samples. (a) Classical fidelity of different sample numbers. When there are 88 simulated samples, $F_c=0.986\pm0.053$. (b) TVD of different sample numbers. When there are 88 simulated samples, $d=0.094\pm0.053$. Error bars are the SEM. Red solid line is the upper limit (lower limit) for the fidelity (TVD).}
	\label{Fc_TVD}
\end{figure}

In Fig. \ref{NSBH}(b), the amplitudes of these processes at different lattice depths along the $x$ direction are shown. When $V_x=2.9 E_{rS}$, $V_y=47.0E_{rS}$ and $V_z=45.4E_{rS}$, the values of these processes are summarized in Table \ref{NSBH3}. We can find that at the experimental parameters, the additional terms are $\sim $10 percent $J$. As the depth of the lattice increases, the non-standard processes are suppressed. When $ V_x=10E_{rS}$ or larger, the additional terms are two orders of magnitude smaller than $J$ and $U$. Thus, for a deeper lattice, these terms could be dropped in the standard Bose-Hubbard model (BHM).

\subsection{Comparison with BHM}
We compared the two models in the evaluations of the classical fidelity. First, we calibrate the chemical potential using the non-standard model and then calculate the ideal probability to evaluate the classical fidelity. The procedure is repeated with the standard BHM. The chosen samples and experimental data for the chemical calibrations are identical, and $J$ and $U$ are the same in the  NSBHM and the standard one. The former's fidelity is higher than the latter by $\sim 2$ percent, which indicates that the NSBHM capturing the additional processes is more suitable in the regime of experimental parameters.

\subsection{Finite-sampling effect}
In the main text, we measure the classical fidelity $F_c$ and the total variance distance $d$ of the output probabilities for an $L=4,N_b=2$ system. The measurement results are $F_c=0.981\pm0.053$ and $d=0.153\pm0.053$, with $88$ samples in total.

Fig. \ref{Fc_TVD} reveals the finite-sampling effect on classical fidelity and TVD \cite{Spring2013}. We draw samples from the ideal probability distribution by a Monte-Carlo method without any noise. 
However, the classical fidelity is below 99\% if the number of samples is less than 100.
This means that the measured results in the main text are underestimated due to the finite-sampling effect.

\begin{figure}[!tbp] 
	\centering
	\includegraphics[width=0.47\textwidth]{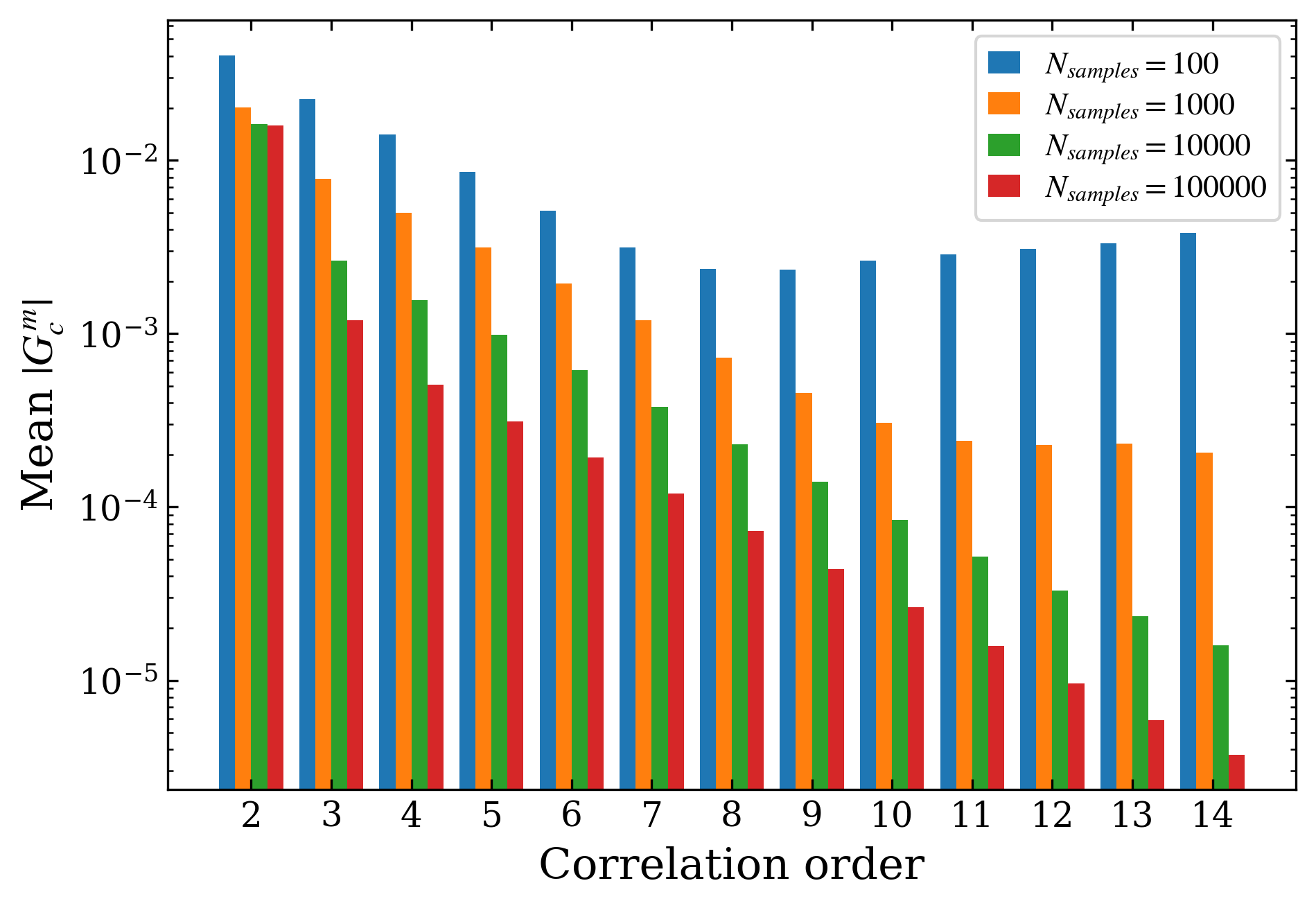}
	\caption{Mean $\left| G_c^m \right|$ for simulated samples from TDVP results in the thermalized phase of a chain with 32 sites and 20 bosons. When the number of samples is added, this abnormal increasing trend in higher orders gradually disappears.}
	\label{FiniteSampleCorr}
\end{figure}

The similar effect was seen while measuring correlations. We extract correlations from about 100 samples for different system sizes, and find an abnormal increase in the higher orders. In Ref.\cite{Rispoli2019a}, this is a kind of quantum critical behavior at the many-body localization transition. But in our work the disorder strength is away from this critical regime both in the thermalized and MBL phases. We attributed this increasing trend to the finite sampling too. We can verify this by classical numerical simulation. When the number of samples is increased, this abnormal phenomenon gradually disappears (Fig.\ref{FiniteSampleCorr}). This rules out the criticality in our experiments. Besides, even with the presence of the finite-sampling effect, the correlations is significantly different in the thermal and MBL phases.

\section{Theoretical evidence of quantum advantage signatures in sampling from the driven thermalized quantum systems}

In this section, we provide a brief theoretical overview on signatures of a sampling quantum advantage in driven thermalized quantum many-body systems. For more details, we refer readers to Ref.~\cite{Thanasilp2020}.

We start by providing formal evidence that sampling output-strings from a periodic evolution of an unitary instance drawn from COE cannot be achieved efficiently using classical computers unless polynomial hierarchy (PH) collapses. The PH non-collapse is a strongly-held conjecture in computer science community with the most famous example of $\text{P} \neq \text{NP}$. 
Then, we argue that thanks to the close connection between COE and an ensemble of Floquet unitaries in the driven thermalized phase, this suggests the same sampling complexity could be expected in the driven thermalized quantum systems.

\subsection{Computational complexity of sampling from COE dynamics}

We first quickly introduce the standard procedure used to demonstrate a sampling quantum advantage in most proposals such as random quantum circuits~\cite{Arute2019}. 
Intuitively, all proposals of a sampling quantum advantage aim to generate quantum evolution that has an sufficient amount of randomness such that classical computers have no structure in the evolution to be exploited and are required to simulate the entire Hilbert space, which is widely believed to be inefficient.

Given protocols to generate a random physical unitary $\hat{U}$, our first step is to assume that there exists a classical machine $\mathcal{C}$ that can efficiently sample from the evolution generated by $\hat{U}$. If we are able to show that the unitaries constructed from the protocols satisfy \textit{the anti-concentration} condition, the computational power of $\mathcal{C}$ can be boosted using the Stock-Mayer theorem to \textit{approximate} the output probability of the quantum evolution. Physically-speaking the anti-concentration implies that most of bit-strings have finite probabilities to be measured. 
Our upgraded classical machine now with the computational power to not only sample from the output distribution but also approximate the output distribution is said to reside in the third level of Polynomial Hierarchy. Our next step is to show that this task of approximating output distribution is \#P-hard (preferably on average). Then, this would imply that our upgraded machine in the third level has an ability to efficiently solve a \#P-hard task, leading to a collapse of Polynomial Hierarchy to the third level. Since it is strongly believed from the computational complexity perspective that the PH collapse cannot happen, the only way to rescue ourselves in this situation is to conclude that the classical machine $\mathcal{C}$ does not exist in the first place.

\begin{figure*}[tbp] 
	\centering
	\includegraphics[width=0.95\textwidth]{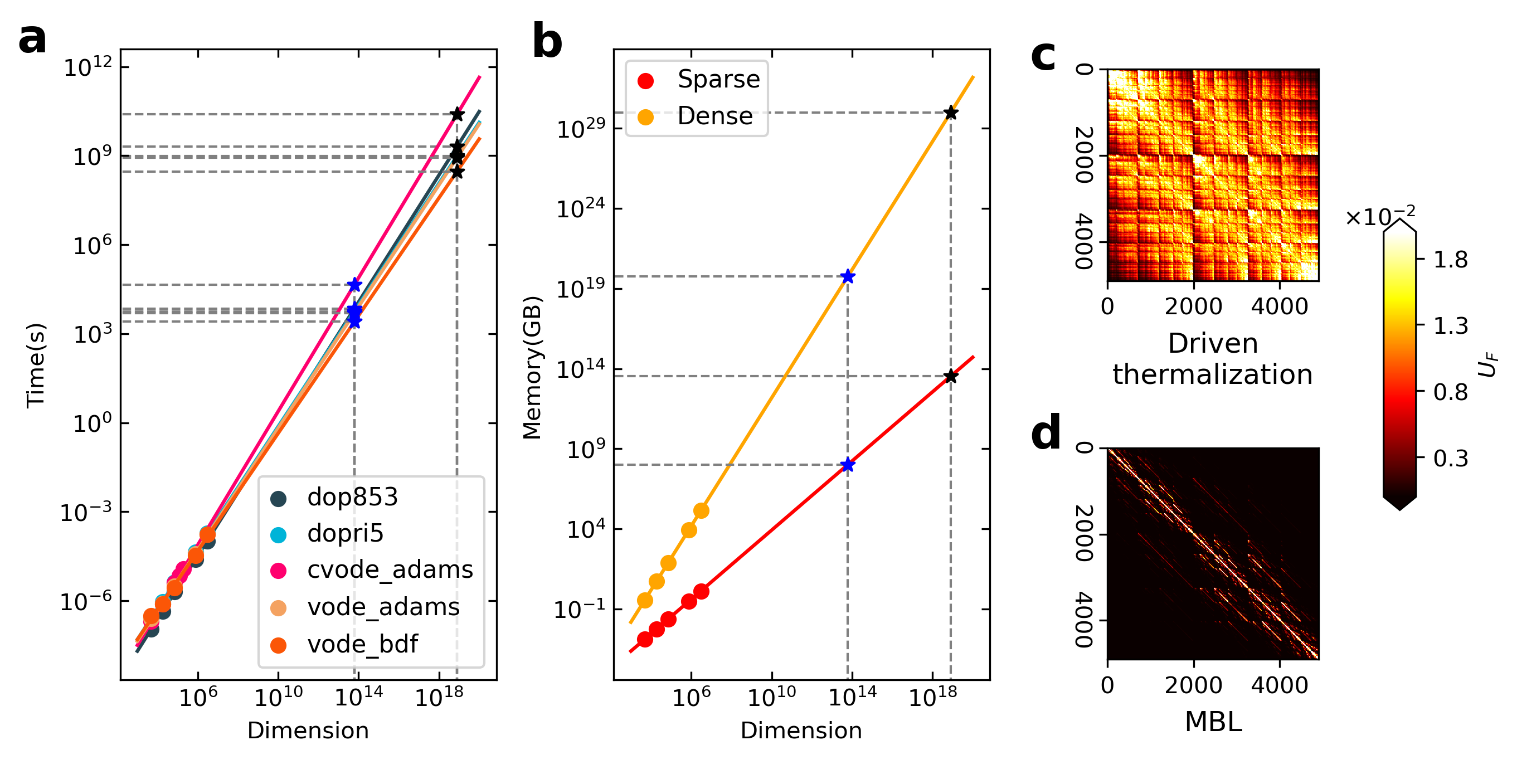}
	\caption{Cost of the classical algorithm. (a), The estimated time cost for the classical simulation algorithm to solve the time-dependent Schr\"{o}dinger equation on the \textit{Frontier} supercomputer (dots). We compare several best-known ODE solvers. In larger system sizes, the \textit{vode\_bdf} method performs best in terms of time cost. The solid lines are linear fitting for logarithms of dimension and time.  The blue and black stars represent the estimates of the largest 1D system and ladder system, respectively. (b), The space complexity for varied system sizes. The orange and red dots denote the consumed storage memory required for sparse (non-zero elements) and dense (full elements) Hamiltonian matrices, respectively. The solid lines are linear fitting for logarithms of dimension and memory. (c) and (d) show the heat maps of a typical Floquet operator in terms of matrix element in the driven thermalized and MBL phases, respectively.}
	\label{ClassicalCost}
\end{figure*}

As one can see, there are two key conditions to show before claiming a quantum advantage in sampling tasks. (i) multiplicative approximation of the output distribution is \#P hard on average and (ii) anti-concentration of the output distribution. Ref~\cite{Thanasilp2020} shows that indeed COE satisfies both conditions. The \#P hardness is proven in the worst case scenario. Particularly, as IQP circuits are a part of COE family, this shows that there exists at least one instance in COE that is \#P hard. 
The average case hardness is conjectured with some theoretical support by mapping from COE dynamics to complex Ising spin-chains.
The anti-concentration is proven using the statistical properties of COE, showing the output distribution satisfies the Porter Thomas distribution.

\subsection{Implication to the driven thermalized systems.}
Now, we discuss the consequences of our results obtained in the previous section for periodically driven quantum many-body systems in the thermalized phase. 
Particularly, COE and the ensemble of Floquet unitaries generated by the driven thermalized quantum systems are closely connected thanks to the external drive which increases the level of randomness in the physical systems.
To understand how the external drive plays a role, we first consider the case of generic \textit{undriven} thermalized systems. Here the quantum system thermalizes to finite temperature due to energy conservation. One can apply random matrix theory to accurately describe systems over narrow energy windows distant from the energy spectrum's boundaries. If the complete energy spectrum is analyzed, the local structure generally observed in static Hamiltonians emerges and random matrix theory is no longer valid. In addition, there are additional constraints due to local few-body interactions. 

This is in contrast to the driven thermalized systems where random matrix theory can be accurately applied to the entire $\hat U_F$ spectrum. Furthermore, under the condition that COE and the driven thermalized systems share the same statistical distributions as indicated by the Floquet eigenstate thermalization hypothesis (ETH). Lastly, using Magnus expansion one can shows that the Floquet Hamiltonian $H_F$ has effective infinite-range multi-body interactions generated by the periodic drive. Consequently, the majority of restrictions from local interactions typically seen in physical systems are removed. 
Therefore, one can speculate the sampling complexity from the generic driven thermalized quantum systems to be the same as sampling from COE.

\begin{figure}[!tbp] 
	\centering
	\includegraphics[width=0.49\textwidth]{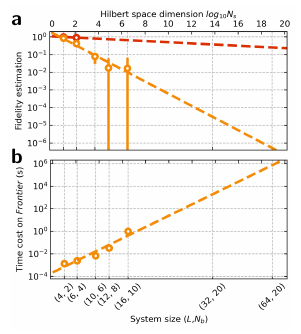}
	\caption{(a), Fidelity estimations of the sampling. The red line is obtained from the classical fidelity $F_c$ while orange line is obtained from the modified cross entropy benchmark $F_d$. Error bars denote the SEM. (b), Taking the estimated fidelity from $F_d$ into account, we fit the classical computation time cost on \textit{Frontier}.}
	\label{fig:Fidelity_fc_fd}
\end{figure}

\section{Classical Algorithms}
\subsection{Schr\"odinger evolution}
To solve the time-dependent Schr\"odinger equation, we adopt an open-source software package, \textit{Quspin}, to compute the evolution of the driven system. Ref. \cite{weinberg2019quspin} provides a detailed introduction to this package. The original version of the package only involves the ordinary differential equation (ODE) solvers provided by \textit{SciPy} \cite{Virtanen2020}. We further added the \textit{cvode} solvers from the \textit{Pyodesys} \cite{Dahlgren2018} to \textit{Quspin}.

In Fig. \ref{ClassicalCost}(a), we plot the required time to solve the time-dependent Schr\"{o}dinger equation for different system sizes. We first record the time spent on \textit{Hanhai20} clusters (with Rpeak 2.38 PFlop/s) at USTC, then we estimate how long it will take on \textit{Frontier} (with Rpeak 1,685.65 PFlop/s) by comparing floating-point computing power, assuming a perfect parallel framework of the code for large scale simulations. The dots of different colors represent the time of systems within the computational capability of the \textit{Hanhai20} clusters. We fit the curves and extrapolate them into the classically intractable regime to determine the required time of our largest 1D system of $ L=32$, $N_b=20 $ (blue stars) and 2D system of $ L=64$, $N_b=20 $(black stars). For 1D system, when the dimension of the Hilbert space reaches $ \sim10^{14} $, the corresponding time to obtain one valid sample is $ \sim2,500 $ s on \textit{Frontier} employing the fastest solver of \textit{vode\_bdf}. For ladder system, the largest dimension is $\sim 10^{19}$, and it will take $\sim 2.9\times 10^8$ s to get a sample, while it takes only 500 s in our experiment.

Fig. \ref{ClassicalCost}(b) illustrates the allocated memory for storing the matrix elements of the sparse (red dots) and dense (orange dots) Hamiltonians. Here the sparse matrix denotes that the matrix contains only non-zero elements, while the dense matrix contains the full elements. 
To store the sparse matrix of the Hamiltonian with double float data type, it requires at least $10^3$ and $10^7$ PB memory for the largest 1D system and ladder system, respectively, while storing the dense matrix requires $10^{13}$ and $10^{23}$ PB memory. Both of them exceed the storage capability of 9.2 PB RAM on the \textit{Frontier}.

\begin{figure}[!tbp] 
	\centering
	\includegraphics[width=0.49\textwidth]{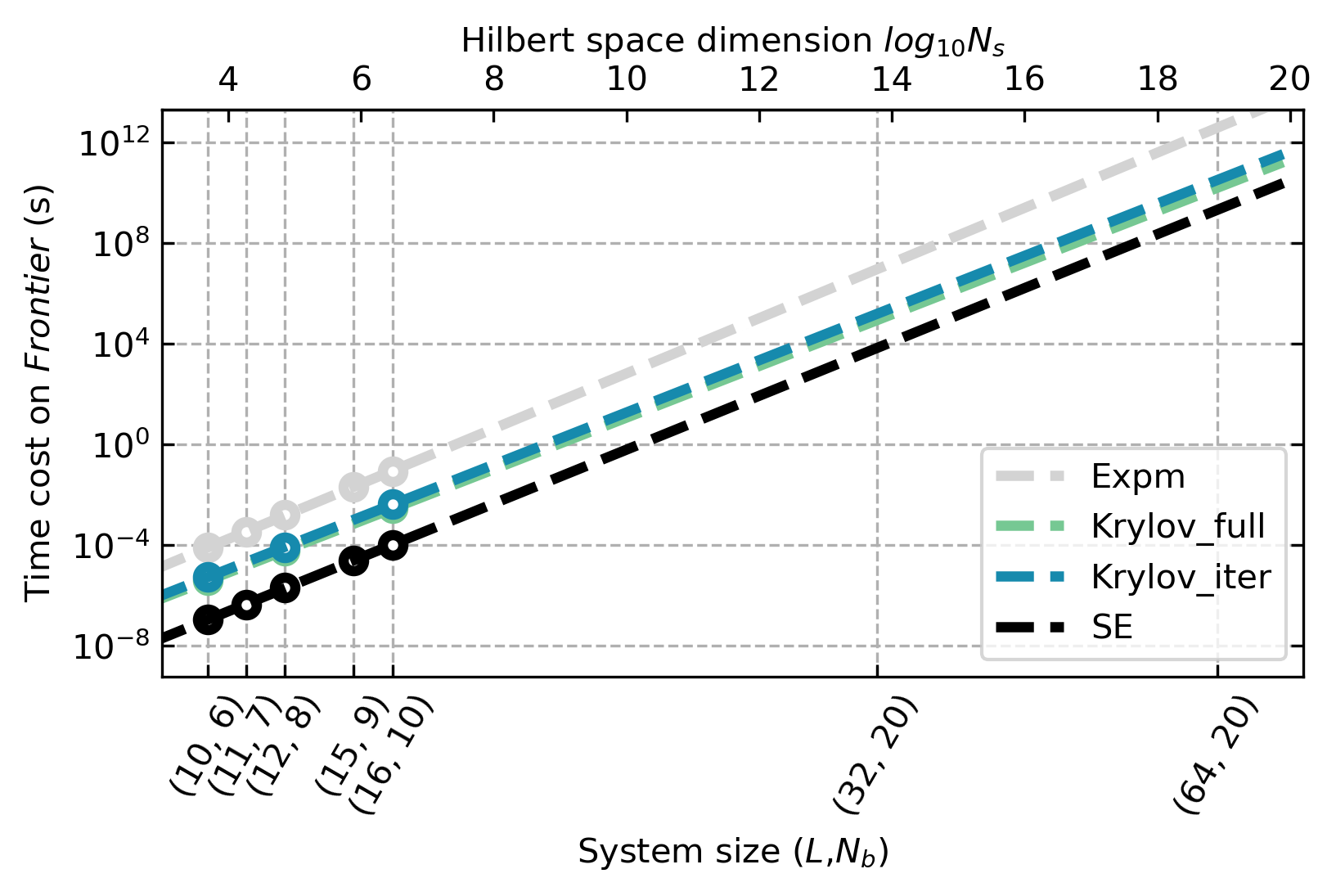}
	\caption{We also tried Krylov subspace as an approximation method to reproduce the sampling task and estimated the time cost time cost on \textit{Frontier}.The gray circles and curve indicates the time cost of time evolution with exact diagonalization of the full matrix of the Hamiltonian which serves as a baseline in comparison with those Krylov subspace methods.}
	\label{fig:Krylov_cost}
\end{figure}

\begin{figure*}[!htbp]
	\centering
	\includegraphics[width=0.95\textwidth]{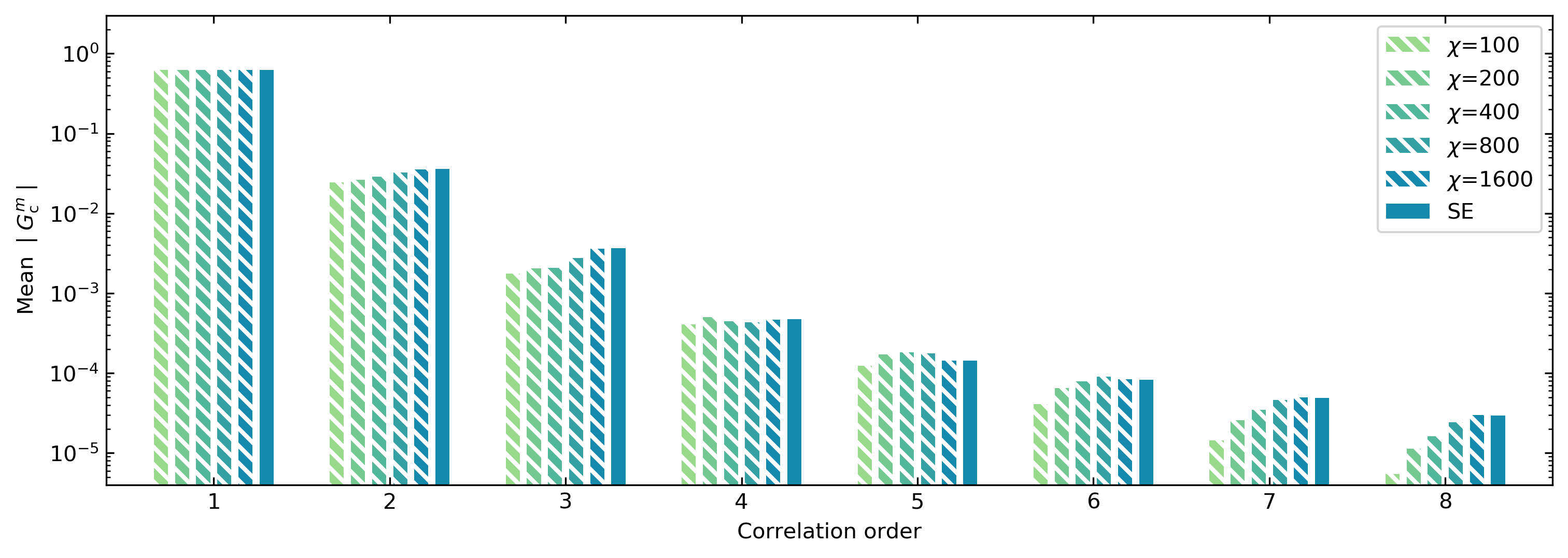}
	\caption{Comparison of numerics. When increasing the bond dimension $\chi$ of the TDVP simulation, the correlation functions approach the values predicted by SE. These calculations are performed for the $L=16$ and $N_b=10$ system. The bond dimension of 1,600 approaches the maximum value of the bond during the evolution.}
	\label{ComNum}
\end{figure*}
\begin{figure*}[!tbp]
	\centering
	\includegraphics[width=0.95\textwidth]{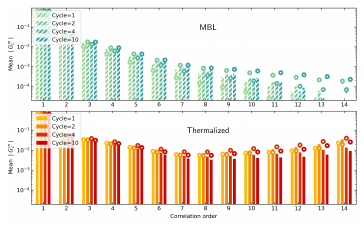}
	\caption{Breakdown of the approximating numerical simulation for long time dynamics in thermalized phase. In the top (bottom) panel we plot the multi-point correlations in the MBL (thermalized) phase. Open circles (bars) denote the correlations extracted from the experimental data (numerical simulation). We measure the correlations for increasing driving cycles. The number of samples used in the analysis is set to 65 for all the settings to remove finite sampling effects. Error bars denote the SEM.}
	\label{fig:Corcycle}
\end{figure*}

\subsection{Time-dependent variational principle}
For large systems, the memory is insufficient in the \textit{Hanhai20} clusters to perform the exact simulation.
Thus we exploit the TDVP \cite{haegeman2016unifying} to simulate the dynamics, which is memory-efficient and is capable of performing the simulation up to $ 64 $ sites. The simulation is implemented based on the package \textit{TeNPy} \cite{hauschild2018efficient}. The time cost of the simulation shows an exponential scaling versus the bond dimension. We fix the bond dimension to 200 for the self-consistent iteration of the local energy offsets mentioned in Section \ref{chemit}, which perfectly approximates the results from SE in terms of the density profile.
In all classical simulations, we restrict the state per site (sps) to be 5.

\subsection{Krylov subspace method}
Instead of exploiting the full matrix of the Hamiltonian to do the time evolution, one can restrict the matrix size within a certain number to the computational cost as the diagonalization is only for the subspace of the full matrix. This approximation is termed as Krylov subspace method and will introduce error depending on the size of the subspace, which is similar with truncation errors in MPS based algorithms like TDVP. In this work we utilize the builtin algorithms in \textit{Quspin}.

\subsection{Fidelity estimations}
Apart from the aforementioned classical fidelity $F_c$, we also adapted one modified cross entropy benchmark $F_d$ \cite{Mark2023,Shaw2024} which is defined as:
\begin{equation}
    \label{equ:Fidelity_fd}
F_d=2\frac{\frac{1}{M}\sum_{m=1}^{M}p(z_m)/p_{avg}(z_m)}{\sum_z p(z)^2/p_{avg}(z)}-1,
\end{equation}
where $p(z)$ is the probability of the bitstring $z$, $p_{avg}(z)$ is the average of $p(z)$ over time. Here we take 1000 points during the 10 cycle driving for averaging. Fig. \ref{fig:Fidelity_fc_fd} (a) shows the estimated fidelities from both methods. 

Given the fidelities extracted from the experimental samples or from the extrapolation at each size, we can now estimate the computation costs of various approximation classical simulations taking these infidelities into account. 
Krylov method still require more time than the SE approach in the regime of this work, even taking the lower fidelities given by $F_d$ into account (Fig. \ref{fig:Krylov_cost}). This makes sense as Krylov method is an approximation method based on exact diagonalization with a time cost scaling $\mathcal{O}(N^2)$ while SE method scales $\mathcal{O}(N)$. 
However, when we tried TDVP, though in the small sizes the time cost is bigger than SE it grows slower than SE (Fig. \ref{fig:Fidelity_fc_fd} (b). In the end, the extrapolated computational cost is $\sim 7\times 10^5$ s for (64,20) system on \textit{Frontier}. In terms of core-hours, the lower bound for classical computation is about 1.7 trillion. The infidelity deducts the quantum speedup of our quantum machine by two orders of magnitude, which still offers a better performance compared with other peer noisy intermediate-scale quantum devices like \textit{Sycamore} \cite{Arute2019} or \textit{Jiuzhang} \cite{Zhong2020}.

\subsection{Comparison of the numerics}
The prediction of the multi-point correlation functions for the (32,20) and (64,20) systems reported in Fig. \ref{fig:cor} is from a TDVP simulation. We found that the predicted values are underestimated compared with the experimental data, especially in higher-order correlations. This is attributed to the cut-off of the bond dimension, which leads to an MPS involving less entanglement. To examine the guess, we vary the bond dimension from 100 to 1,600 for the simulations in the (16,10) system. As shown in Fig. \ref{ComNum}, for smaller bond dimensions, the lower-order correlations are consistent with those from SE, while the higher-order correlations are weaker than those from SE. Until the bond dimension increases to the upper limit, i.e., there is no cut-off during the evolution, the eighth-order correlation approaches the value predicted by SE. For larger systems and higher-order correlations, the required bond dimension of the TDVP simulation to reproduce the experimental results will increase as well, which is infeasible even with the TDVP method.

\begin{figure*}[!tbp] 
	\centering
	\includegraphics[width=0.9\textwidth]{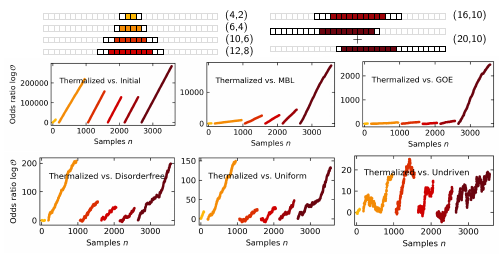}
	\caption{Odds ratios. We plot the logarithm of the odds ratio $\mathcal{O}(H_0,H_1|S_n)$ versus number of samples. As the magnitude of the ratio itself is too huge we take the logarithm of the odds ratio to the base 10. Legends are defined the same as Fig.\ref{fig:BayandCon}.}
	\label{fig:odds}
\end{figure*}

\section{Data analysis}
\subsection{Sample collection}
The average filling of the initial state is about 96\%. Besides, some of the atoms will hop outside the chain during the driving dynamics or expanding for detection. Thus, the realizations in which there are atoms outside the chain are rejected. We also post-select the samples that the total number of the detected atoms in the $L$-site chain is exact $N_b$ and the atom number in each site is less than 5. Similarly, in the entanglement entropy measurements, we post-select the realizations where the total parity in the 2$\times$6-site plaquette is even. For the largest system of 32 sites and 20 atoms, we typically retain 8\% of the data, corresponding to a sampling rate of one sample every 500 seconds.

\subsection{Mock-up distributions}

We propose six competitive mock-up distributions in the Bayesian tests. The Initial mock-up is that the probability of the string corresponding to the initial state is unity, and is zero otherwise.

\begin{equation}
p_{\mathrm{init}}(z_i)=\left \{
\begin{aligned}
    &1, & z_i=01\cdots10,\\
    &0, & \mathrm{otherwise},
\end{aligned}
\right.
\end{equation}

The second mock-up is the Uniform distribution, which is a flat distribution over the whole string.

\begin{equation}
    p_{\mathrm{uni}}(z_i)=\frac{1}{N_s},
\end{equation}
where $N_s$ is the dimension of the Hilbert space. 
The other mock-ups are derived from the NSBHM, which share identical Hubbard parameters except one of them with the ideal probability distribution. For instance, the Disorderless mock-up is calculated based on a NSBHM without disorder potential, and the Undriven one is derived from that without the driving term. For the MBL and GOE mock-up, we choose an amplitude of 20$\times W$ for the disorder potential and a driving frequency of 20$\times \omega$ for the simulations, respectively. All the calculations are based on SE.

\subsection{Bayesian hypothesis tests and odds ratios}
To validate the samples efficiently, the Bayesian tests are proposed in Ref. \cite{Bentivegna2014} and later were used in a series work of boson sampling experiments \cite{Bentivegna2014,Wang2017c,Zhong2021a,Madsen2022}. In this section, we give a brief introduction of the test. For further details, please refer to this article \cite{Bentivegna2014}.

Given $n$ samples $S_n=\left[s_1,s_2,\cdots,s_n\right]$from the quantum machine, we need decide from which sampler those samples are. One hypothesis $H_0$ is the samples are generated from the ideal sampler and the other $H_1$ is that from the mockup sampler. According to Bayes' rule \cite{Pishro-Nik2014}, the posterior probabilities of $H_0$ and $H_1$  are 

\begin{equation}
\label{equ:Baysth}
\begin{split}
    P(H_0|S_n)=\frac{P(S_n|H_0) P(H_0)}{P(S_n)}, \\
    P(H_1|S_n)=\frac{P(S_n|H_1) P(H_1)}{P(S_n)},
\end{split}
\end{equation}
where $P(H_0)$ and $P(H_1)$ are prior probabilities satisfying $P(H_0)+P(H_1)=1$, $P(S_n|H_0)=\prod_{i=1}^n P_{\mathrm{ideal}}(s_i)$ and $P(S_n|H_1)=\prod_{i=1}^n P_{\mathrm{mk}}(s_i)$ are the conditional probabilities for the two hypotheses, respectively. Then, we can compare those posterior probabilities and choose the hypothesis with a higher value. That is , we choose $H_0$ when $P(H_0|S_n)\ge P(H_1|S_n)$.  The ratio between them

\begin{equation}
\label{equ:oddsratio}
    \mathcal{O}(H_0,H_1|S_n)=\frac{P(H_0|S_n)}{ P(H_1|S_n)}=\frac{P(S_n|H_0) P(H_0)}{P(S_n|H_1) P(H_1)},
\end{equation}
is the so-called odds ratio. Without loss of generality, we assume the even prior probabilities $P(H_0)=P(H_1)=1/2$ due to the lack of any prior information. Now the odds ratio reads $\mathcal{O}(H_0,H_1|S_n)=P(S_n|H_0)/P(S_n|H_1)$. The Bayesian likelyhood defined in the main text is thus related to the odds ratio

\begin{equation}
\label{equ:baytoodd}
\begin{split}
    \mathcal{B}(\mathrm{ideal},\mathrm{mk})&=\frac{P(S_n|H_0)}{P(S_n|H_0) +P(S_n|H_1)}\\
    &=\frac{\mathcal{O}(H_0,H_1|S_n)}{\mathcal{O}(H_0,H_1|S_n)+1}.
    \end{split}
\end{equation}
So if the Bayesian likelyhood $\mathcal{B}(\mathrm{ideal},\mathrm{mk})\geq 0.5$, or equivalently, the odds ratio $\mathcal{O}(H_0,H_1|S_n)\ge 1$, we choose the hypothesis of $H_0$, which means the samples are generated from the ideal sampler rejecting the mock-up sampler.
Similarly, the confidence of the Bayesian test shown in Fig. \ref{fig:BayandCon} b could be derived from the odds ratio as well. In Fig. \ref{fig:odds} we plot the accumulated odds ratio versus number of samples.
\begin{equation}
\label{equ:bayconfodd}
\begin{split}
    \Delta \mathcal{B} (\mathrm{ideal},\mathrm{mk})_{n} &= \frac{1}{n}\log \frac{P(S_n|H_0)}{P(S_n|H_1)}\\
    &=\frac{1}{n}\log \mathcal{O}(H_0,H_1|S_n).
    \end{split}
\end{equation}

\subsection{Multi-point correlation functions}
The joint expectation value of operators $ \{ \hat{n}_i\} $ is $\langle \prod_{i=1}^m \hat{n}_i \rangle=\langle \hat{n}_1 \hat{n}_2 \dots \hat{n}_m\rangle $, captures two kinds of correlations: ``disconnected'' correlations and ``connected'' correlations. 
The multi-point density correlation function measured in our experiment is the ``connected'' part. It includes only the $ m $-th order correlations and cannot be described by lower-order correlations.

In the main text, Eq. \ref{equ:Cor} gives the recursive formula of the multi-point density correlation function. Here, we write down a four-point case (Eq. \ref{equ:Corr4}) explicitly as an example:

\begin{widetext} 
    \begin{align} 
        G_{\mathrm{c}}^4(x_1,x_2,x_3,x_4)
        =&G_{\mathrm{tot}}^4(x_1,x_2,x_3,x_4)-G_{\mathrm{dis}}^4(x_1,x_2,x_3,x_4)  
        \nonumber  \\
        =&\langle \hat{n}_1 \hat{n}_2 \hat{n}_3 \hat{n}_4\rangle - \Big[ G_{\mathrm{c}}^3(x_1,x_2,x_3)\langle \hat{n}_4 \rangle + G_{\mathrm{c}}^3(x_1,x_2,x_4)\langle \hat{n}_3 \rangle + G_{\mathrm{c}}^3(x_1,x_3,x_4)\langle \hat{n}_2 \rangle + G_{\mathrm{c}}^3(x_2,x_3,x_4)\langle \hat{n}_1 \rangle \Big]
        \nonumber  \\
        &- \Big[ G_{\mathrm{c}}^2(x_1,x_2) G_{\mathrm{c}}^2(x_3,x_4) + G_{\mathrm{c}}^2(x_1,x_3) G_{\mathrm{c}}^2(x_2,x_4) + G_{\mathrm{c}}^2(x_1,x_4) G_{\mathrm{c}}^2(x_2,x_3) \Big] 
        \nonumber  \\
        &- \Big[ G_{\mathrm{c}}^2(x_1,x_2) \langle \hat{n}_3 \rangle \langle \hat{n}_4 \rangle + G_{\mathrm{c}}^2(x_1,x_3) \langle \hat{n}_2 \rangle \langle \hat{n}_4 \rangle + G_{\mathrm{c}}^2(x_1,x_4) \langle \hat{n}_2 \rangle \langle \hat{n}_3 \rangle + G_{\mathrm{c}}^2(x_2,x_3) \langle \hat{n}_1 \rangle \langle \hat{n}_4 \rangle 
        \nonumber  \\
        &+ G_{\mathrm{c}}^2(x_2,x_4) \langle \hat{n}_1 \rangle \langle \hat{n}_3 \rangle + G_{\mathrm{c}}^2(x_3,x_4) \langle \hat{n}_1 \rangle \langle \hat{n}_2 \rangle \Big] - \langle \hat{n}_1 \rangle \langle \hat{n}_2 \rangle \langle \hat{n}_3 \rangle \langle \hat{n}_4 \rangle
    \label{equ:Corr4}
    \end{align}
\end{widetext}

We find that the ``disconnected'' part consists of the integer partitions of the involved lattice sites with at least two nonzero integers. For the three-point case, the groupings are $\{ (2,1),(1,1,1) \}$, where $(2,1)$ contains three cases: $(x_1,x_2;x_3)$, $(x_1,x_3;x_2)$ and $(x_2,x_3;x_1)$. Hence, $G_{\mathrm{dis}}^3$ has four terms in total. For the four-point case, as mentioned above, we can divide them into $\{ (3,1),(2,2),(2,1,1),(1,1,1,1) \}$, and the number of cases for each partition is $\{ 4, 3, 6, 1\}$, leading to a total amount of fourteen terms in $G_{\mathrm{dis}}^4$.

\begin{figure}[!tbp] 
	\centering
	\includegraphics[width=0.47\textwidth]{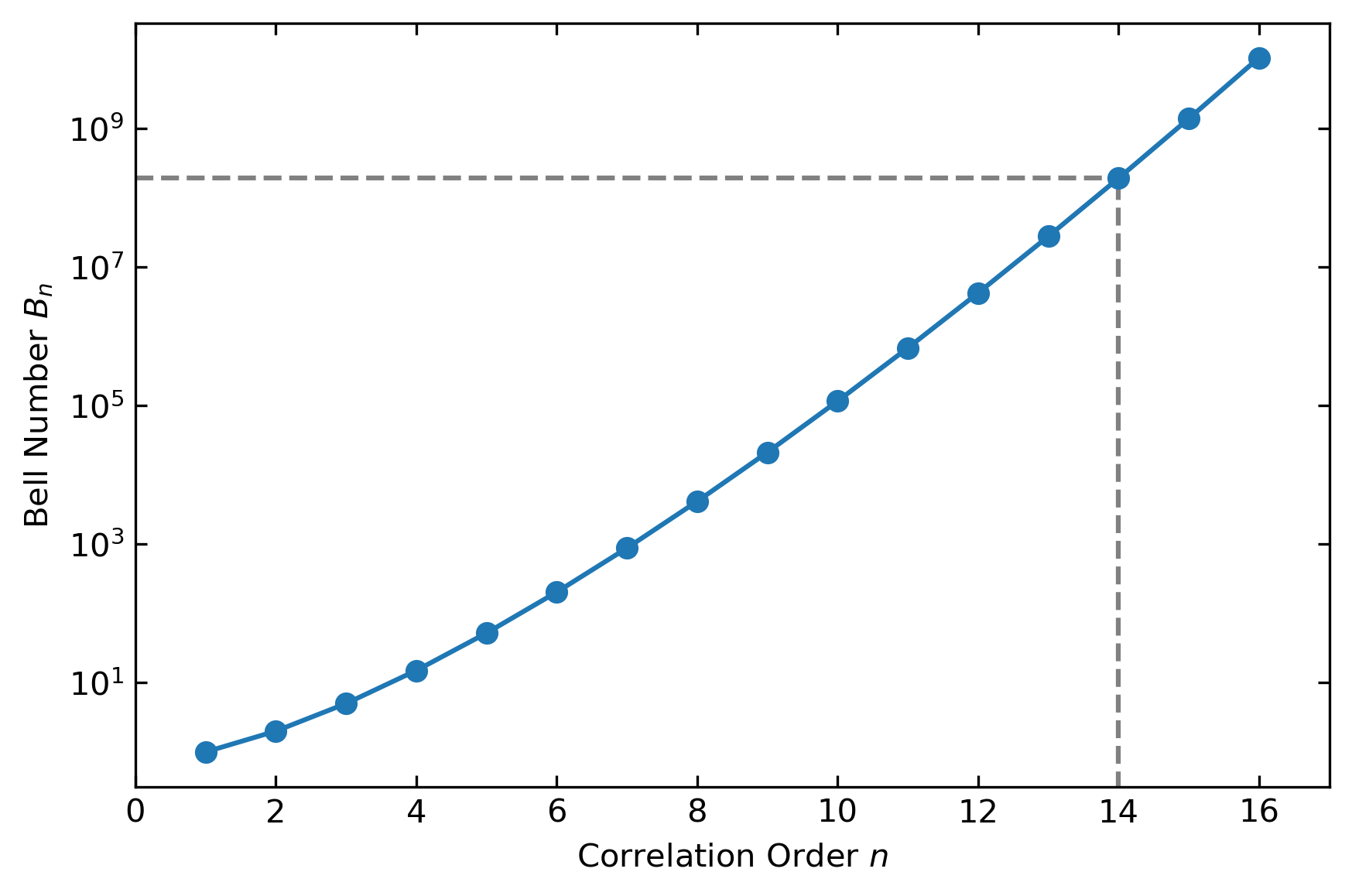}
	\caption{Bell Numbers. 
    The number of terms in multi-point density correlation functions shows an explosive growth trend as the order increases. The highest order we calculated for $L=32,N_b=20$ is $14$th, which contains about $10^8$ lower order terms.}
	\label{Bell_Numbers}
\end{figure}

\begin{figure}[!tbp] 
	\centering
	\includegraphics[width=0.47\textwidth]{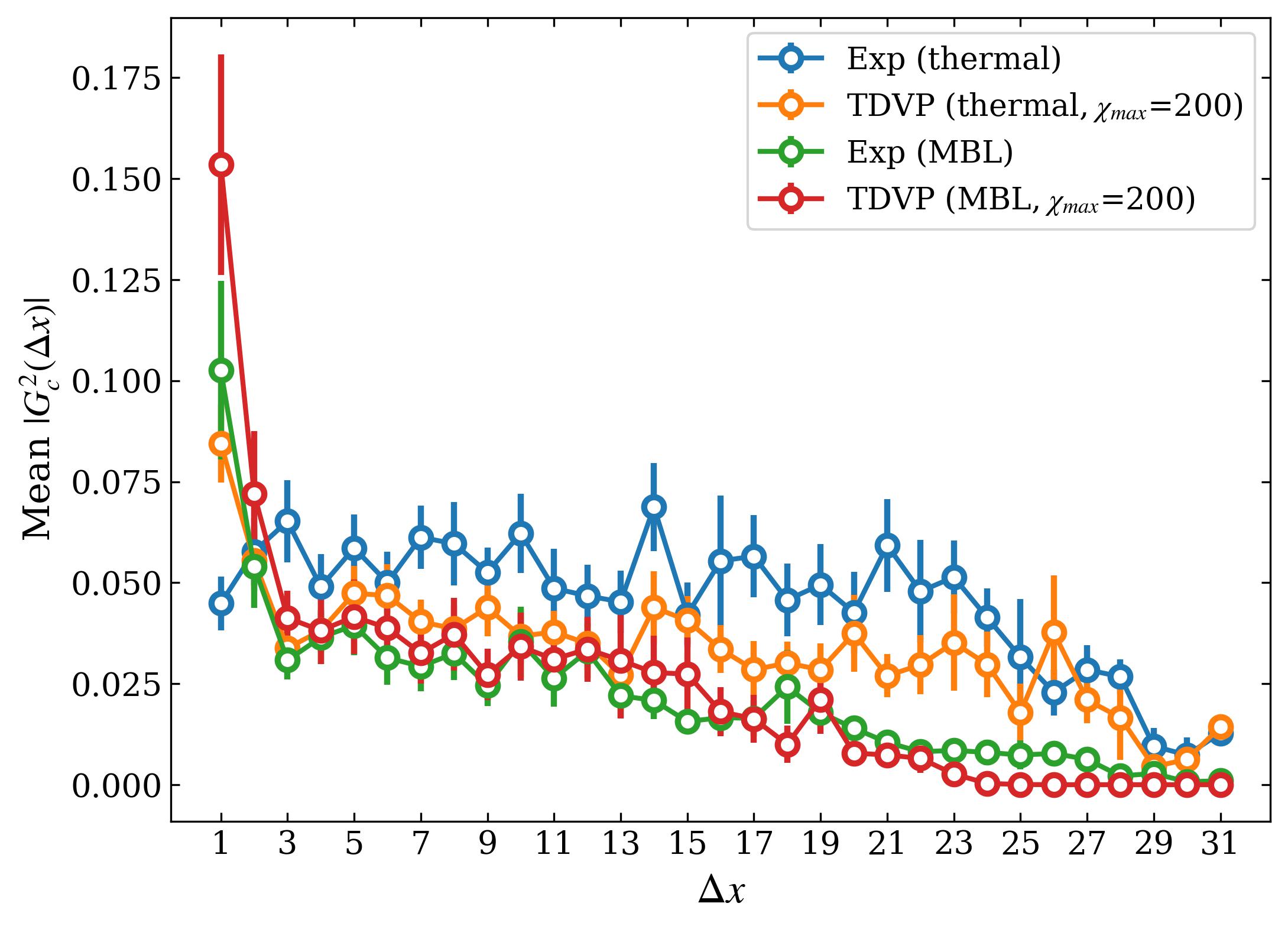}
	\caption{Two-point correlations. 
    We plot the connected two-point correlations over the distance between the two points. 
    The system size is (32,20), and the number of driving cycles is 10. 
    Blue (Green) points are experimental data in the thermalized (MBL) phase. Orange (Red) points are predictions from TDVP simulations for the thermalized (MBL) phase. Error bars denote the SEM.}
	\label{fig:cortwopoints}
\end{figure}
The number of terms in the ``connected'' correlation ($G_{\mathrm{tot}}^m$ is also included here), is called the Bell numbers, corresponding to a partitioning problem. In Ref. \cite{bell1934exponential}, E.T. Bell gave the definition of Bell numbers: The Bell number $ B_n $ is the number of ways to divide a set with $n$ elements into disjoint, nonempty subsets. $B_0=1$ because there is only one partition of an empty set. Eq. \ref{equ:Bell_rf} reveals the recursive relationship of the Bell numbers in terms of binomial coefficients (Ref. \cite{wilf2005generatingfunctionology}).
\begin{equation} \label{equ:Bell_rf}
B_{n+1}=\sum_{k=0}^n{C_n^k B_k}
\end{equation}
Therefore, as the order of the correlation increases, the number of terms in multi-point density correlation functions will have an explosive growth trend, which is shown in Fig. \ref{Bell_Numbers}. The $14$th-order correlation contains about $10^8$ lower-order terms. Besides, to obtain the mean value of $|G_{\mathrm{c}}^{14}(\mathbf{x})|$, there are $C_{32}^{14}$ configurations of subsystems involving 14 sites in the 32-site chain.
Hence, in the system of (32,20), we select only 100 configurations randomly to average.

\subsubsection{Two-point correlation functions}
In Fig.~\ref{fig:cortwopoints}, we plot the 2-point connected correlator for a $L=32$,$N_b$=20 system with 10-cycle driving in both thermalized and MBL phases. 
First, in the thermalized phase, the correlator calculated from numerical simulation (orange line, TDVP method with $\chi_{max}=200$) drops rapidly with the distance between two sites $\Delta x$ increasing, while that from experimental data (blue line) holds large values over a long distance $\Delta x$. 
For experimental data, the curve is almost flat, which implies long-range correlations across the whole system. 
The drop in the tail is attributed to the boundary effects. Second, in the MBL phase, the correlation length is shorter, but we can still see comparable values in the two-point correlations of distance $\Delta x \le $10 sites. 
As in short distances, atoms can still hop around to build up correlations. 
It also indicates that we need to go high-order correlations to distinguish the thermalized and MBL phases.

\subsubsection{Comparison of the multi-point correlations between driven and undriven thermalized phases}
\begin{figure}[!tbp] 
	\centering
	\includegraphics[width=0.47\textwidth]{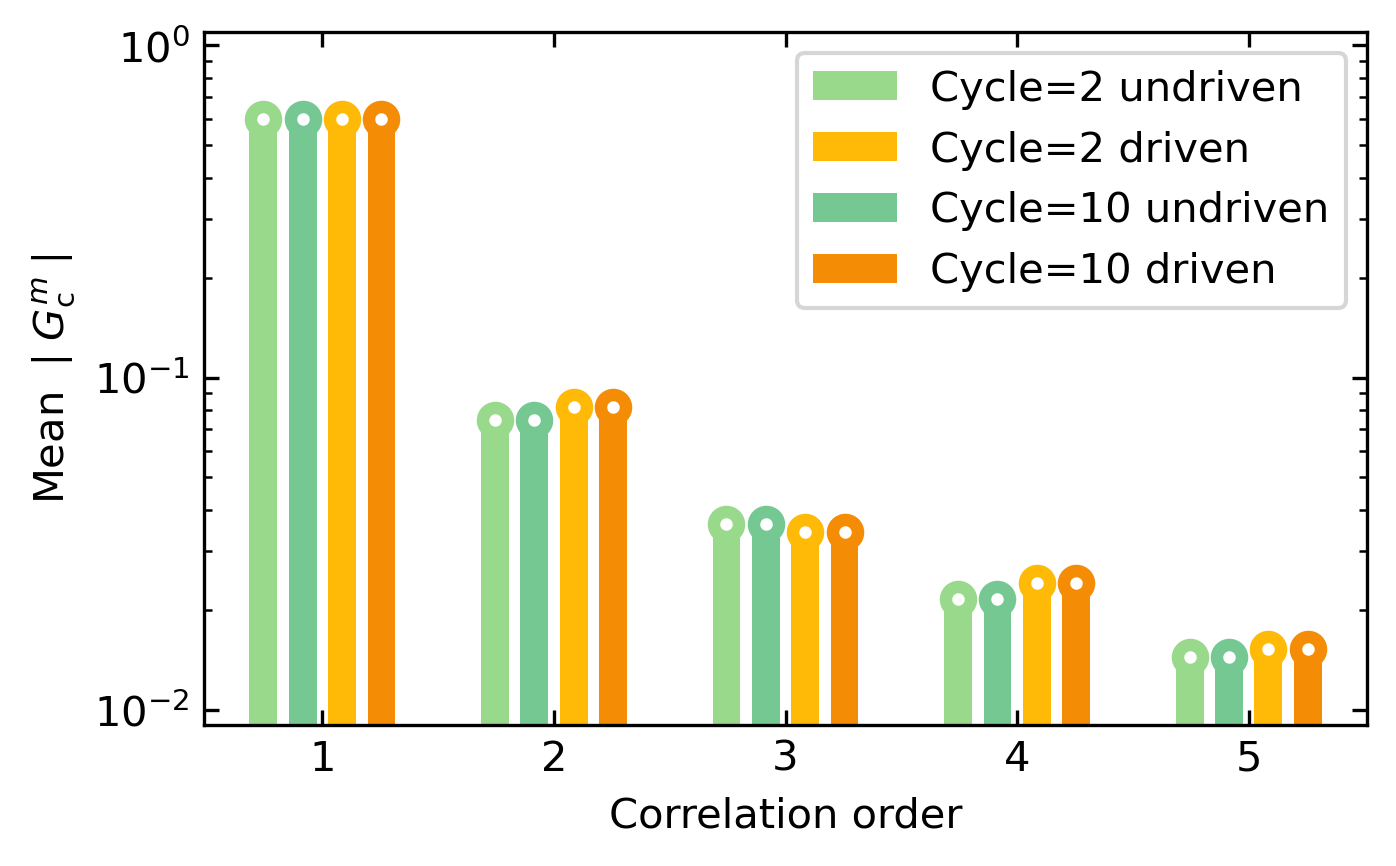}
	\caption{Comparison of the multi-point correlations between driven and undriven thermalized phases. 
    We plot the connected multi-point correlations up to 5th order in a system of (10,6). We fixed the number of samples in each case to 51. Open circles are experimental data and error bars indicate the SEM. The bars are predictions from the numerical simulations of SE.}
	\label{fig:cor_undriven}
\end{figure}

To further rule out possible classical heating to the driven thermalized phase, we checked the correlations over various time holding when switching off the driving. As shown in Fig. \ref{fig:cor_undriven}, there is no noticeable difference in the multi-point correlations for longer holding time without driving. This is suggesting that there is negligible classical heating injected to the system up to 10 cycles.

\FloatBarrier
\renewcommand{\addcontentsline}[3]{}

\bibliographystyle{naturemag}

\end{document}